\documentclass[11pt]{article}

\usepackage[margin=1in]{geometry}
\usepackage{setspace}
\usepackage{natbib}
\usepackage{graphicx}
\usepackage{amsmath,amssymb}
\usepackage{bm}
\usepackage{siunitx}
\usepackage{adjustbox}
\usepackage{multirow}
\usepackage{float}
\usepackage{authblk}
\usepackage{placeins}
\usepackage{hyperref}
\usepackage{cleveref}    
\usepackage{amsthm}
\usepackage{enumitem}
\usepackage{caption}
\usepackage{booktabs}

\DeclareMathOperator*{\argmax}{arg\,max}
\DeclareMathOperator{\Var}{Var}
\DeclareMathOperator{\Cov}{Cov}

\newcommand{\E}{\mathbb{E}}
\newcommand{\R}{\mathbb{R}}

\newtheorem{assumption}{Assumption}
\newtheorem{theorem}{Theorem}
\newtheorem{lemma}{Lemma}
\newenvironment{condition}[1]
  {\par\medskip\noindent\textbf{Condition #1.}\itshape}
  {\medskip}

\title{Nonparametric Proportional Hazards Model with Differential Regularization Applied to Spatial Survival Data}

\author[1]{Lorenzo Tedesco\thanks{lorenzo.tedesco@unibg.it --- Equally contributing author.}}
\author[1]{Francesco Finazzi\thanks{Equally contributing author.}}
\affil[1]{Department of Economics, University of Bergamo, Via dei Caniana 2, 24127 Bergamo (BG), Italy}

\date{} 

\hypersetup{
  pdftitle={Nonparametric Proportional Hazards Model with Differential Regularization},
  pdfauthor={L. Tedesco, F. Finazzi},
  colorlinks=true,
  linkcolor=blue,
  citecolor=blue,
  urlcolor=blue
}

\begin{document}

\maketitle

\begin{abstract}

The Proportional Hazards (PH) model is one of the most widely used models in survival analysis, typically assuming a log-linear relationship between covariates and the hazard function. However, in the context of spatial survival data, where the time-to-event variable is associated with a spatial location within a given domain, this assumption is often unrealistic in capturing spatial effects. Thus, this paper proposes modeling the location effect through a nonparametric function of spatial location. The function is approximated using finite element methods on a triangulated mesh to accommodate irregular domains. Estimation is carried out within the classical partial likelihood framework, with smoothness of the spatial effect enforced through differential penalization. Using sieve methods, we establish the consistency and asymptotic normality of the parametric component. Simulations and two empirical applications demonstrate superior performance compared to existing approaches. 
\end{abstract}

\section{Introduction}\label{sec:introduction}

Since its introduction by \citep{Cox:1972}, the proportional hazards (PH) model has become a fundamental tool in survival analysis. Central to this framework is the hazard function, which quantifies the instantaneous risk of an event occurring at a particular time, given that the subject has not yet experienced the event up to that point.  

Within the PH framework, the hazard is expressed as the product of two parts: a baseline hazard, which is the same for all individuals, and a term that incorporates the effects of covariates. The standard version of the model assumes that the covariates have a log-linear effect, meaning the logarithm of the hazard is a linear combination of the covariates. This setup provides a convenient interpretation of the regression coefficients as hazard ratios: each coefficient reflects the multiplicative change in risk associated with a one-unit increase in the corresponding covariate. These coefficients can be estimated efficiently using partial likelihood, which focuses on the ordering of event times rather than their exact distribution. This approach has the important advantage of profiling out the baseline hazard, allowing it to remain unspecified while still yielding consistent and efficient estimates of the covariate effects. As a result, inference can be conducted on hazard ratios without making parametric assumptions about the underlying hazard function.  

The assumption of log-linearity, however, can be too restrictive in practice. A more flexible alternative is the nonparametric proportional hazards model, which allows some covariate effects to be modeled in a nonparametric way rather than forcing a purely linear form. In this formulation, the hazard depends on two sets of covariates: one that enters the model linearly, with a finite set of regression coefficients, and another that enters through an unspecified smooth function. 

Estimation approaches for the PH model that incorporate nonparametric covariate effects through smooth functions fall into two main categories. The first class consists of local likelihood methods \citep{tibshirani1987local, fan1997local, chen2010global}, which rely on kernel smoothing and the careful selection of bandwidth parameters. The second widely used class consists of spline-based approaches \citep{o1988nonparametric, hastie1990exploring, kooperberg1995hazard}, in which the smooth function is approximated using a basis expansion. This formulation enables flexible modeling while preserving parsimony and interpretability.

In this paper, we focus on the case where the nonparametric smooth function reflects a spatial effect in a bidimensional domain. In such contexts, standard nonparametric approaches fail to adequately represent spatial variation in the hazard. In fact, both local likelihood methods and spline-based approaches are primarily designed for univariate smoothing, and their direct extensions to two-dimensional spatial domains often suffer from instability, boundary bias, or prohibitive computational cost. Consequently, they may fail to capture complex spatial structures in the hazard function, highlighting the need for a spatially-aware nonparametric extensions of the PH model.

For a comprehensive review of spatial PH models, we refer the reader to \cite{hanson2014spatial}, and we restrict our discussion to the main contributions. In \cite{li2006semiparametric}, the authors propose a semiparametric normal transformation model for spatial survival data. In this framework, observations marginally follow a PH model, while their joint distribution is defined by transforming the data into approximately normal variables and assuming a multivariate normal distribution for the transformed outcomes. A key limitation of this approach is that, if the transformation is misspecified, the assumption of multivariate normality may be violated, potentially leading to biased inference. Notably, this model was originally motivated by the claim that ``direct nonparametric maximum likelihood estimation in such models is practically prohibited due to the high-dimensional, intractable integration in the likelihood function and the infinite-dimensional nuisance baseline hazard parameter.'' This highlights the importance of basing the estimation on the partial likelihood, which, as in the standard PH model, treats the infinite-dimensional baseline hazard as a nuisance parameter, rather than attempting estimation from the full likelihood. Following this principle, the proposed model assumes independence among observations, incorporates a nonparametric spatial effect into the hazard function to capture spatial effects, and employs the partial likelihood for estimation, thereby showing that a nonparametric estimation can be carried out in a straightforward manner.

Another spatial PH alternative is the composite likelihood approach of \cite{paik2012composite}, which assumes the Farlie--Gumbel--Morgenstern distribution and models the dependence parameter as a function of geographic and demographic pairwise distances. Apart from the restrictive dependence structure, composite likelihood methods---while computationally convenient---can be less efficient than full likelihood approaches, as they ignore higher-order dependencies.  

Bayesian formulations have also been proposed, such as \cite{banerjee2003frailty,taylor2017}, which partition the domain into clusters and assign random effects.
In \cite{hennerfeind2003geoadditive}, instead, a nonparametric spatial effect is modeled using splines, where spatial dependence is incorporated through the choice of priors on the spline parameters. These methods, however, rely on a parametric specification of the baseline hazard, thereby overlooking a key advantage of leaving the baseline hazard as a nuisance parameter.

Finally, a limitation common to all the cited spatial PH models is the assumption that spatial dependence is driven solely by Euclidean distance, thereby neglecting the geometry of the domain. This simplification can lead to biased results in settings with irregular boundaries, non-convex shapes, or internal holes.


In this work, we introduce a spatial PH model that remains within the classical partial likelihood framework. We model a continuous, nonparametric spatial effect over a two-dimensional domain, enabling a smooth representation of spatial variation in the hazard. The method is implemented via finite element methods (FEM) on a mesh constructed from a triangulation of the domain of interest, ensuring accurate representation even for highly irregular domains. Spatial variation in survival data can be modeled through geostatistical approaches, which rely on continuous coordinates (e.g., latitude and longitude), or through lattice-based approaches, which model dependence among discrete spatial units. Our method can accommodate both, and while focusing on geostatistical data, we also describe how it naturally extends to areal data, see Section~\ref{sec:areal} of the Appendix.  

The smoothness of the spatial effect is enforced through a differential penalization term, which yields a concave maximization problem for a high-dimensional, differentiable objective function. This formulation enables efficient estimation using derivative-based optimization methods, facilitated by the fact that the derivative is available in closed form and simple to evaluate. A related penalization strategy for spatial regression is discussed in \cite{sangalli2013spatial}, although it cannot be readily extended to survival data. 
Another distinction is that our estimation procedure is developed within the framework of sieve theory, which enables the derivation of robust asymptotic properties. Thus, our approach could also serve as a further step toward achieving asymptotic results in spatial regression with differential regularization.

The paper is organized as follows. In Section~\ref{sec:model} we specify the model. In Section~\ref{sec:estimation} we develop the estimation procedure based on penalized likelihood and the sieve method with finite element approximations. Section~\ref{sec:asymptotic} establishes the asymptotic properties of the proposed estimator. Section~\ref{sec:simulations} presents simulation results illustrating finite-sample performance, while Section~\ref{sec:empirical} provides empirical applications. Section~\ref{sec:conclusion} contains our concluding remarks.


\section{Model}\label{sec:model}
Let $T$ be a nonnegative time-to-event random variable representing the occurrence of the event of interest. We consider a compact spatial domain with non-empty interior $\Omega \subset \mathbb{R}^2$ with a regular boundary $\partial \Omega \in C^2(\mathbb{R}^2)$. 
Let $\bm{X} \in \mathbb{R}^b$ denote a $b$-dimensional vector of covariates, and let $\bm{P}$ be a vector taking values in $\Omega$, representing the spatial location at which an observation is made. 
We assume that, conditional on the values $\bm{X} = \bm {x}$ and $\bm{P} = \bm{p}$, the hazard function of $T$ follows the PH model
\begin{align}\label{eq:model}
	\lambda(t \mid \bm{x}, \bm{p})
	= \lambda_0(t) \exp\!\big( \bm{x}^\top \bm{\beta}_0 + h_0(\bm{p}) \big),
\end{align}
where $\lambda_0(t)$ is an unspecified baseline hazard function shared across all individuals; $\boldsymbol{\beta}_0 \in \mathbb{R}^b$ is a vector of regression coefficients that quantifies the log-linear effects of the covariates $\boldsymbol{X}$ on the hazard; and $h_0 : \Omega \to \mathbb{R}$ is an unknown smooth function that captures residual spatial variation in the hazard associated with location $\boldsymbol{P}$, that is not explained by the observed covariates.

As in the standard interpretation of the proportional hazards model, the coefficient $\beta_{0j}$ represents the log hazard ratio associated with a one-unit increase in the $j$-th covariate, holding all other covariates and the spatial effect constant. The spatial term $h_0(\bm{p})$ serves to identify regions of elevated or reduced hazard relative to the baseline $\lambda_0(t)$. Since $h_0$ is identifiable only up to an additive constant, we impose the centering condition
\begin{equation}\label{eq:identification_h}
\int_{\Omega} h_0(\bm{p})\,\mathrm{d}\bm{p} = 0,
\end{equation}
which guarantees uniqueness and will prove to be computationally convenient.

Because of right-censoring, it is not possible to directly observe $T$. Instead, we observe
\[
Y = \min(T, C), 
\quad \delta = I(T \le C),
\]
where $C$ denotes a censoring variable and $I(\cdot)$ is the indicator function. We consider the following assumption.

\begin{assumption}\label{assumption:identifiability}
	(i) Conditional on $\bm{X}$ and $\bm{P}$, the censoring variable $C$ is independent of $T$.
	(ii) There exists $\tau > 0$ such that $\mathbb{P}(T \ge \tau) > 0$ and $\mathbb{P}(C \ge \tau) > 0$, ensuring that the support of $Y$ is non-degenerate. $(iii)$ The true coefficient vector \(\bm\beta_0\) is an interior point of a compact set \(\mathcal B \subset \mathbb{R}^b\). $(iv)$ The matrix \(\mathbb E[\bm X\bm X^\top]\) is positive definite. $(v)$ The spatial effect $h_0$ is smooth and satisfies \eqref{eq:identification_h}. 
\end{assumption}
Assumption~\ref{assumption:identifiability}~$(i)$ is standard in survival analysis and ensures unbiased estimation under right-censoring. 
Assumption~\ref{assumption:identifiability}~$(ii)$ guarantees non-degenerate support of the observed data, ensuring sufficient follow-up for reliable inference. 
Assumption~\ref{assumption:identifiability}~$(iii)$ imposes compactness of the parameter space, a technical condition that facilitates consistency of the estimator. 
Assumption~\ref{assumption:identifiability}~$(iv)$ requires positive definiteness of $\mathbb{E}[\bm X \bm X^\top]$, preventing collinearity and ensuring identifiability of regression effects. 
Assumption~\ref{assumption:identifiability}~$(v)$ enforces smoothness and centering of the spatial effect, which provide identifiability and regularity for asymptotic analysis. 

\begin{theorem}
	\label{theo:identification}
Suppose Assumption~\ref{assumption:identifiability} holds. Then, the model in~\eqref{eq:model} is uniquely identified.
\end{theorem}
	The proof is reported in Appendix~\ref{appendix:identification}.

 \section{Estimation via Penalized Likelihood and Sieve Method}\label{sec:estimation}
In order to characterise the regression structure and capture spatial heterogeneity, we have developed an estimation procedure based on penalised likelihood. Specifically, we assume that the observed data arise from an independent and identically distributed sample of $(Y, \delta, \bm X, \bm P)$ of size $n$, namely 
\(
\{(Y_i, \delta_i, \bm{x}_i, \mathbf{p}_i)\}_{i=1}^n.
\)  
Our objective is to estimate the regression coefficients \(\boldsymbol{\beta}_0\) and the spatial effect \(h_0\) by maximizing a penalized log-partial-likelihood functional $Q_n$ defined on a suitable space $\mathcal{B}\times\mathcal{H}$ as
	\begin{equation}\label{eq:penalized_likelihood}
		Q_n(\bm \beta, h) = \frac{1}{n} \sum_{i=1}^n \delta_i \Big(\bm{x}_i^\top \bm{\beta} + h(\bm{p}_i) - \log \frac{1}{n} \sum_{j=1}^n I(Y_j \ge Y_i) \exp (\bm{x}_j^\top \bm{\beta} + h(\bm{p}_j) ) \Big) - \frac{\lambda}{2} \int_{\Omega} (\Delta h(\bm p))^2 \,\text{d}\bm p.
	\end{equation}
	Here, $\lambda$ is a positive smoothing parameter (potentially dependent on $n$), and $\Delta$ is the Laplacian operator, $\Delta h = \frac{\partial^2 h}{\partial p_1^2} + \frac{\partial^2 h}{\partial p_2^2}$, which measures the local curvature of the effect $h$. The penalty term controls the smoothness of the estimated function $h$, with a larger $\lambda$ enforcing a smoother function.
	
The functional space $\mathcal{H}$ must be chosen carefully. Since the penalty term $\int_{\Omega} (\Delta h)^2$ must be well-defined, we require that $\mathcal{H} \subset H^2(\Omega)$, the Sobolev space of functions in $L^2(\Omega)$ with all distributional derivatives up to order 2 also in $L^2(\Omega)$. By the Sobolev embedding theorem, $H^2(\Omega) \subset C^0(\Omega)$, ensuring that any $h \in H^2(\Omega)$ is continuous and can be evaluated pointwise at any location. To facilitate the estimation procedure, we consider homogeneous Neumann boundary conditions, meaning the flux across the boundary is zero: $\nabla h \cdot \mathbf{n} = 0 \text{ on } \partial \Omega$, where $\mathbf{n}$ is the outward-pointing normal unit vector. Combining these requirements with the identifiability constraint, we consider the following functional parameter space:
	$$
	\mathcal{H} = \left\{ h \in H^2_{\mathbf{n}}(\Omega) \;\middle|\; \int_{\Omega} h \, dx = 0, \|h\|_{H^2(\Omega)} < M_{\mathcal{H}} \right\},
	$$
	where $H^2_{\mathbf{n}}(\Omega) = \left\{ h \in H^2(\Omega) \;\middle|\; \nabla h \cdot \mathbf{n} = 0 \text{ on } \partial \Omega \right\}$, and $M_{\mathcal{H}}$ is a constant large enough. The bound $\|h\|_{H^2(\Omega)}<M_{\mathcal{H}}$ is a consequence of the smoothness of $h_0$ and the fact the $\Omega$ is compact. 
	
Maximizing the functional in~\eqref{eq:penalized_likelihood} is an infinite-dimensional optimization problem. To make it computationally tractable, we replace the infinite-dimensional space $\mathcal{H}$ with a sequence of finite-dimensional subspaces $\mathcal{H}_n$ that become dense in $\mathcal{H}$ as $n \to \infty$. This is the core idea of the method of sieves \citep{chen2007large}. We use the finite element method to construct these subspaces.

\subsection{The Finite Element Method}
The FEM is a versatile numerical technique for approximating differential operators. Its core idea is to construct a finite-dimensional approximation space, typically spanned by locally supported basis functions. These basis functions are defined with respect to a subdivision of the computational domain $\Omega$ into simpler subdomains \citep{quarteroni2009numerical}.

We consider a triangulation $\mathcal{T}_\eta$ of $\Omega$, in which the domain is partitioned into non-overlapping triangles such that any two adjacent triangles share either a complete edge or a single vertex. The parameter $\eta$ denotes the mesh size, typically taken as the diameter of the largest triangle in the mesh. In this setting, the triangulation also serves to approximate the boundary $\partial\Omega$ by a polygon (or, more generally, a union of polygonal segments), ensuring that even curved boundaries are represented in a piecewise linear fashion.  

\paragraph{Illustration with $C^0$ linear elements.}
From this triangulation, one often constructs a finite-dimensional function space spanned by locally supported basis functions. In the case of continuous, piecewise linear finite elements ($C^0$), each basis function $\psi_k$ is associated with a specific vertex $\bm{\xi}_k$ of the triangulation. By definition, $\psi_k$ takes the value $1$ at its corresponding vertex $\bm{\xi}_k$ and $0$ at all other vertices, i.e.,
\[
\psi_k(\bm{\xi}_j) = \delta_{kj},
\]
where $\delta_{kj}$ is the Kronecker delta. Any function $h$ in this $C^0$ finite element space can be expressed as
\[
h(\bm{p}) = \sum_{k=1}^{K} c_k \, \psi_k(\bm{p}) = \mathbf{c}^\top \bm{\psi}(\bm{p}),
\]
where $K$ is the total number of vertices, and the coefficients $c_k$ correspond to nodal values, $c_k = h(\bm{\xi}_k)$.

\begin{figure}[h]
	\centering
	\includegraphics[width=0.6\textwidth]{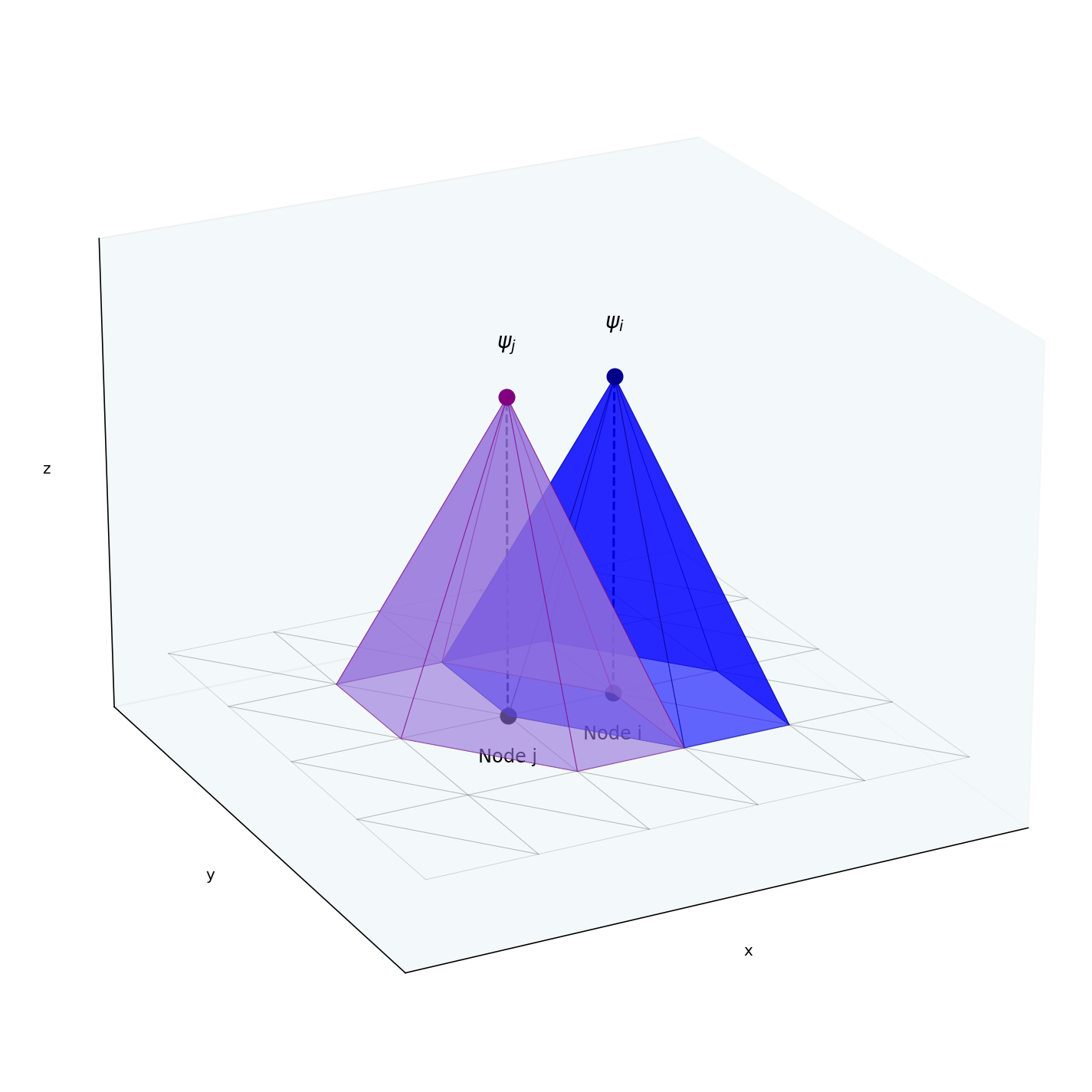}
	\caption{Example of a linear finite element basis function on a planar mesh. The function is locally supported, taking the value one at a specific vertex and zero at all other vertices. 
    }
	\label{fig:linear_basis_function}
\end{figure}

\paragraph{$C^1$ space used in our estimator.}
While the $C^0$ FEM construction presented above is helpful for intuition, our penalization involves the Laplacian-squared term $\int_\Omega (\Delta h)^2$, which is only well-defined for $h\in H^2(\Omega)$. Therefore, for estimation we employ an $H^2$-conforming, $C^1$ finite element space (e.g., the Argyris element, see \citealp{brenner2008mathematical}). For each triangulation $\mathcal{T}_\eta$ we define
\[
V_\eta = \Big\{\, v \in C^1(\overline{\Omega}) \;\Big|\; v|_T \in \mathbb P_5(T)\ \text{for every } T \in \mathcal{T}_\eta \,\Big\},
\]
where $\mathbb P_5(T)$ denotes the space of bivariate polynomials on the triangle $T$ of total degree $\le 5$ (dimension $21$). The associated degrees of freedom (DOFs) enforce $C^1$ continuity across elements (values, first and second derivatives at vertices, and edge-midpoint normal derivatives), ensuring $V_\eta \subset H^2(\Omega)$.

Let $\{\psi_k\}_{k=1}^{K(\eta)}$ be the global $C^1$ basis associated with these DOFs, defined by Kronecker interpolation with respect to the DOF functionals. Any $h\in V_\eta$ can be written as
\[
h(\bm p) \;=\; \sum_{k=1}^{K(\eta)} c_k\,\psi_k(\bm p) \;=\; \mathbf c^\top \bm\psi(\bm p),
\]
where now $\mathbf c$ collects DOF values (not just nodal values).

\subsection{Sieve Space Construction}\label{sec:sieve_space_construction}
To apply sieve theory, we define a sequence of triangulations and corresponding function spaces. Let $\{\eta(n)\}_{n \in \mathbb{N}} \subset (0,1]$ be a decreasing sequence converging to zero. For each $\eta=\eta(n)$, we have a conforming triangulation $\mathcal{T}_{\eta}$ of the domain $\Omega$ satisfying
\[
\max_{T \in \mathcal{T}_{\eta}} \operatorname{diam}(T) \le \eta \, \operatorname{diam}(\Omega).
\]
This family of triangulations is said to be \emph{quasi-uniform}, meaning there exists a constant $\rho > 0$ such that for all $\eta$,
\[
\min_{T \in \mathcal{T}_{\eta}} \operatorname{diam}(B_T) \ge \rho\, \eta\,  \operatorname{diam}(\Omega),
\]
where $B_T$ is the largest ball contained in triangle $T$. This condition prevents triangles from becoming arbitrarily thin, ensuring good approximation properties.
The discrete space corresponding to $H^2_{\mathbf n}(\Omega)$ is
\[
H_n = V_{\eta(n)} \cap H^2_{\mathbf n}(\Omega).
\]
Finally, the sieve space for our problem is the finite-dimensional approximation of $\mathcal{H}$:
\[
\mathcal H_n = \Big\{\, h \in H_n \;\Big|\; \int_\Omega h\,d\bm p = 0,\ \|h\|_{H^2(\Omega)} < M_{\mathcal H} \,\Big\}.
\]

The quality of this approximation is guaranteed by standard $C^1$ FEM interpolation theory. For any $h \in \mathcal{H}$, we can define a projection $\mathcal{J}^{\,n} : \mathcal{H} \to \mathcal{H}_n$ as
\[
\mathcal J^{\,n} h = \mathcal I^{\,n} h - \frac{1}{|\Omega|} \int_{\Omega} \mathcal I^{\,n} h\, d\bm p,
\]
where $\mathcal I^{\,n}$ is a $C^1$-conforming interpolation operator (e.g., the Argyris interpolant). Then (see, e.g., \citealp{brenner2008mathematical}),
\begin{equation}\label{eq:bound_approx}
	\|h - \mathcal{J}^{\,n} h\|_{\infty} \le C\, \eta(n)\, \|h\|_{H^2(\Omega)},
\end{equation}
for a constant $C$ independent of $h$ and $\eta(n)$. This shows that the approximation error vanishes at a rate $\mathcal{O}(\eta(n))$ in the $L^\infty$-norm, confirming that $\mathcal{H}_n$ is a suitable sieve space.

The estimation problem is now restricted to the sieve space $\Theta_n = \mathcal{B} \times \mathcal{H}_n$. The estimator $\hat \theta_n = (\hat{\bm \beta}_n, \hat h_n)$ of $\theta_0 = (\bm\beta_0,h_0)$ is found by maximizing the penalized sample likelihood $Q_n(\theta)$ over the sieve space $\Theta_n$:
\begin{align}\label{eq:estimator1}
	\hat \theta_n = \argmax_{\theta \in \Theta_n} Q_n(\theta).
\end{align}

\subsection{Numerical Implementation}\label{sec:implementation}

Let $\bm{\psi}_n(\cdot) = (\psi_1(\cdot), \dots, \psi_{K(n)}(\cdot))^\top$ be the vector of $K(n)$ nodal basis functions for the conforming $C^1$ finite element space $V_n$. Any $h \in \mathcal{H}_n$ can be written as
\[
h(\bm p) = \bm{\psi}_n(\bm p)^\top \mathbf c,\quad \mathbf c \in \mathbb R^{K_n}.
\]
Let the mass and stiffness matrices be
\[
(\mathbf R_{0,n})_{ij}=\int_\Omega \psi_i\psi_j\,d\bm p,
\qquad
(\mathbf R_{1,n})_{ij}=\int_\Omega \nabla\psi_i\cdot\nabla\psi_j\,d\bm p.
\]
The zero-mean constraint $\int_\Omega h\,d\bm p=0$ can therefore be written as $\mathbf 1^\top \mathbf R_{0,n}\mathbf c=0$.
Let $\mathbf r_0=\mathbf R_{0,n}^\top\mathbf 1$ and let $\mathbf Z_n\in\mathbb R^{K_n\times(K_n-1)}$ be an orthonormal basis of $\ker(\mathbf r_0^\top)$; then any feasible $\mathbf c$ can be written as $\mathbf c=\mathbf Z_n\mathbf h$, with $\mathbf h\in\mathbb R^{K_n-1}$.

Introduce $g\in V_n$ as the $L^2$–projection of the (weak) Laplacian:
\[
\int_\Omega g\,v\,d\bm p \;=\; \int_\Omega (\Delta h)\,v\,d\bm p
\;=\; -\int_\Omega \nabla h\cdot\nabla v\,d\bm p
\quad\forall v\in V_n,
\]
where the boundary term vanishes by the homogeneous Neumann condition.  
Writing $g(\bm p)=\bm{\psi}_n(\bm p)^\top \mathbf d$ yields the linear relation
\[
\mathbf R_{0,n}\,\mathbf d \;=\; -\,\mathbf R_{1,n}\,\mathbf c
\quad\Longrightarrow\quad
\mathbf d \;=\; -\,\mathbf R_{0,n}^{-1}\mathbf R_{1,n}\,\mathbf c.
\]
Hence the Laplacian–squared penalty becomes
\[
\int_\Omega (\Delta h)^2\,d\bm p \;\approx\; \|g\|_{L^2(\Omega)}^2
= \mathbf d^\top \mathbf R_{0,n}\mathbf d
= \mathbf c^\top \mathbf R_{1,n}\mathbf R_{0,n}^{-1}\mathbf R_{1,n}\mathbf c.
\]

After reparameterization $\mathbf c=\mathbf Z_n\mathbf h$, the estimator in \eqref{eq:estimator1} is obtained by maximizing
\begin{align}\label{eq:numeric_operator}
	\hat Q(\bm \beta, \bm h) = \frac{1}{n}\sum_{i=1}^n \delta_i\!\left(
	\bm x_i^\top \bm\beta + \bm{\psi}_n(\bm P_i)^\top \mathbf Z_n \mathbf h
	- \log S_n^{(0)}(\bm\beta,\mathbf h,Y_i)\right)
	\;-\; \frac{\lambda}{2}\,\mathbf h^\top \mathbf A_n \mathbf h,
\end{align}
where
\[
S_{n}^{(0)}(\bm{\beta}, \mathbf{h}, t)
= \frac{1}{n}\sum_{j=1}^n I(Y_j\ge t)\,
\exp\!\left(\bm{x}_j^\top \bm\beta + \bm{\psi}_n(\bm{p}_j)^\top \mathbf Z_n \mathbf h\right)
\]
and
\[
\mathbf A_n \;=\; \mathbf Z_n^\top\,\mathbf R_{1,n}\mathbf R_{0,n}^{-1}\mathbf R_{1,n}\,\mathbf Z_n.
\]
The objective is differentiable and concave in $(\bm\beta,\mathbf h)$, so derivative-based maximization applies. In fact we propose to use BFGS quasi-Newton algorithm \citep{nocedal2006numerical} with analytic gradients. For that,  define
\[
\eta_j = \bm{x}_j^\top \bm\beta + \bm{\psi}_n(\bm{p}_j)^\top \mathbf Z_n \mathbf h,
\qquad 
\mathbf r_j = \mathbf Z_n^\top \bm{\psi}_n(\bm{p}_j).
\]

For \(t \ge 0\), set
\[
S_{n,X}^{(0)}(\bm\beta,\mathbf h,t) = \frac{1}{n}\sum_{j=1}^n I(Y_j \ge t)\,e^{\eta_j}\,\bm{x}_j,
\qquad
S_{n,h}^{(1)}(\bm\beta,\mathbf h,t) = \frac{1}{n}\sum_{j=1}^n I(Y_j \ge t)\,e^{\eta_j}\,\mathbf r_j.
\]

Then the gradient of the objective in \eqref{eq:numeric_operator} is
\[
\nabla_{\bm\beta}\hat Q(\bm \beta, h)
= \frac{1}{n}\sum_{i=1}^n \delta_i \left(
\bm{x}_i - \frac{S_{n,X}^{(1)}(\bm\beta,\mathbf h,Y_i)}{S_n^{(0)}(\bm\beta,\mathbf h,Y_i)}
\right),
\]
\[
\nabla_{\mathbf h}\hat Q(\bm \beta, h)
= \frac{1}{n}\sum_{i=1}^n \delta_i \left(
\mathbf r_i - \frac{S_{n,h}^{(1)}(\bm\beta,\mathbf h,Y_i)}{S_n^{(0)}(\bm\beta,\mathbf h,Y_i)}
\right) \;-\; \lambda\,\mathbf A_n\,\mathbf h.
\]

The proposed estimator can therefore be written as 
\begin{equation}\label{eq:estimator}
\hat{\theta}_n = \left(\hat{\bm \beta}_n = \hat{\bm \beta},\;\;
\hat h_n(\cdot) = \bm{\psi}_n(\cdot)^\top \mathbf Z_n \hat{\mathbf{h}}\right),
\end{equation}
where $(\hat{\bm\beta},\hat{\mathbf{h}}) = \argmax_{(\bm \beta,\mathbf{h}) } \hat Q(\bm \beta,\mathbf{h})$.

\subsection{Selecting the tuning parameter}\label{sec:selection}
For the selection of the tuning parameter $\lambda$, we adopt a data-driven cross-validation strategy, as in \cite{cygu2021pcoxtime}. 
In this approach, the optimal value of $\lambda$ is chosen to either minimize the cross-validated partial likelihood deviance (CV-PLD).  

Using cross-validation, the training data are partitioned into $K$ folds. For each $k=1,\dots,K$, the model is fitted on the retained data (all but fold $k$) to obtain penalized estimates 
$(\hat{\bm\beta}_{-k}(\lambda), \hat f_{-k}(\lambda))$, which are then evaluated on the held-out fold. 
The CV-PLD is defined as
\begin{equation}\label{eq:cvpld}
	\text{CV-PLD}(\lambda)
	= -2 \sum_{k=1}^{K} \Big\{
	\ell_{\text{partial}}\!\big(\hat{\bm\beta}_{-k}(\lambda), \hat h_{-k}(\lambda); \mathcal D\big)
	-
	\ell_{\text{partial}}\!\big(\hat{\bm\beta}_{-k}(\lambda), \hat h_{-k}(\lambda); \mathcal D_{-k}\big)
	\Big\},
\end{equation}
where $\ell_{\text{partial}}(\cdot;\mathcal D)$ denotes the log-partial likelihood evaluated on the full dataset $\mathcal D$, that is
\[
\ell_{\text{partial}}(\bm \beta, h; \mathcal D) 
= \frac{1}{n} \sum_{i=1}^n \delta_i \Bigg(\bm{x}_i^\top \bm{\beta} + h(\bm{P}_i) 
- \log \frac{1}{n} \sum_{j=1}^n I(Y_j \ge Y_i)\exp\{\bm{x}_j^\top \bm{\beta} + h(\bm{P}_j)\}\Bigg),
\]
and $\ell_{\text{partial}}(\cdot;\mathcal D_{-k})$ is the same expression computed on the retained (training) data $\mathcal D_{-k}$. 
Subtracting the two ensures that only the contribution of the held-out fold is isolated while preserving the correct risk sets, which typically improves upon the simpler strategy of evaluating the partial likelihood solely on the held-out fold. 
The optimal value of $\lambda$ is then the minimizer of \eqref{eq:cvpld}.

Alternatively, the CV-C-index \citep{harrell1996multivariable} criterion aggregates the out-of-fold risk scores and computes Harrell’s concordance index for the PH model (as implemented, for example, in the \texttt{survival} package; see \citealp{dai2019cross}). 
In this case, the optimal $\lambda$ maximizes the cross-validated C-index.

\section{Asymptotic Analysis}\label{sec:asymptotic}

We now turn to the asymptotic properties of the sieve estimator $\hat\theta_n$. 
Our approach follows the general framework of sieve M-estimation developed in \cite{chen2007large}. Let $d(\cdot, \cdot)$ denote a metric on the parameter space $\Theta$, defined by
\[
d(\theta,\tilde \theta) = \|\beta-\tilde \beta\| + \|h-\tilde h\|_{\infty},
\]
where $\|\cdot\|$ is the Euclidean norm.  
This metric allows us to measure closeness in both the finite-dimensional and infinite-dimensional components of $\theta = (\beta,h)$.

As is customary in sieve methods, let $K(n)$ denote the dimension of the approximating space $\mathcal{H}_n$. Before presenting the main theorems, we introduce regularity conditions that ensure the sieve estimator is well-behaved and facilitate the asymptotic derivations.

\begin{assumption}\label{assumption:density_P}
	(i) The support $\mathcal{X}$ of the covariates $\bm X$ is bounded, i.e.,
	\(
	\sup_{\bm x\in \mathcal{X}} \|\bm x\|\le M_{\mathcal{X}}.
	\) (ii) The triangulation $\mathcal{T}_\eta$, indexed by $\eta$, is assumed to be quasi-uniform.
\end{assumption}

The first step is to establish consistency of the sieve estimator. The following theorem shows that, under mild growth restrictions on $K(n)$ and the penalty parameter, the estimator converges in probability to the true parameter at a controlled rate.

\begin{theorem}[Consistency]\label{theo:consistency}
	Suppose that  assumption~\ref{assumption:identifiability}--\ref{assumption:density_P} hold,  and that
	\begin{align}\label{eq:growth_rule}
		K(n)\log K(n)=o(n)\quad \text{and}\quad  \lambda_n=o(1)
	\end{align} 
	Define
	\[
	\varepsilon_n \;=\; \max\Big\{\,\delta_n,\ \eta(n),\, \sqrt{\lambda_n}\,\Big\},
	\qquad
	\delta_n \asymp \sqrt{\frac{K(n)\log K(n)}{n}}.
	\]
	Then
	\[
	d(\hat\theta_n,\theta_0) \;=\; O_P(\varepsilon_n).
	\]
\end{theorem}
The proof is reported in Appendix~\ref{appendix:consistency}.  We now show that, with a suitable choice of sieve dimension, the parametric component $\hat{\bm\beta}_n$ is asymptotically normal. 

\begin{theorem}[Asymptotic Distribution]\label{thm:beta-CLT-rewrite}
Suppose assumptions of Theorem~\ref{theo:consistency} holds, and $\varepsilon_n^2 = o(n^{-1/2})$,  $\lambda_n \epsilon_n = o(n^{-1/2})$ and $\eta(n)\epsilon_n  = o(n^{-1/2})$. Then, it holds
	\[
	\sqrt{n}\,\big(\hat{\bm\beta}_n - \bm\beta_0\big)
	\;\Rightarrow\; \mathcal N\!\big(0,\ \Sigma_\beta\big),
	\]
	for a suitable covariance matrix $\Sigma_\beta$ specified in the proof.
\end{theorem}

The proof is reported in Appendix~\ref{appendix:normality}.  
Together, Theorems~\ref{theo:consistency} and~\ref{thm:beta-CLT-rewrite} establish that the sieve estimator is both consistent and asymptotically normal in its finite-dimensional component, providing the basis for inference on $\bm\beta_0$.

Note that we can consider 
\(
\eta(n) \;\asymp\; K(n)^{-1/2}.
\)
In fact, for a quasi-uniform triangulation with mesh parameter $\eta(n)$, 
the number of triangles satisfies 
\(
\#\mathcal{T}_{\eta(n)} \;\sim\; C\, \eta(n)^{-2}.
\)
Since each triangle carries a fixed number of local degrees of freedom 
(21 for the Argyris element) and the global $C^1$-continuity only modifies 
the scaling by a constant factor, the dimension of the FEM space grows like 
\(
K(n) \;=\; \dim V_{\eta(n)} \;\asymp\; \eta(n)^{-2}.
\)
Thus, it is easy to check that the condition rates of Theorem~\ref{thm:beta-CLT-rewrite} are satisfied by taking $K(n) = n^{\alpha}$ and $\lambda_n = n^{-\gamma}$ for $\frac{1}{8}<\alpha <\frac{1}{2}$ and $\gamma >\frac{1}{2}$.

\section{Simulation Results}\label{sec:simulations}

\begin{figure}[htp]
    \centering

    \begin{minipage}[t]{0.48\textwidth}
        \centering
        \includegraphics[width=\textwidth]{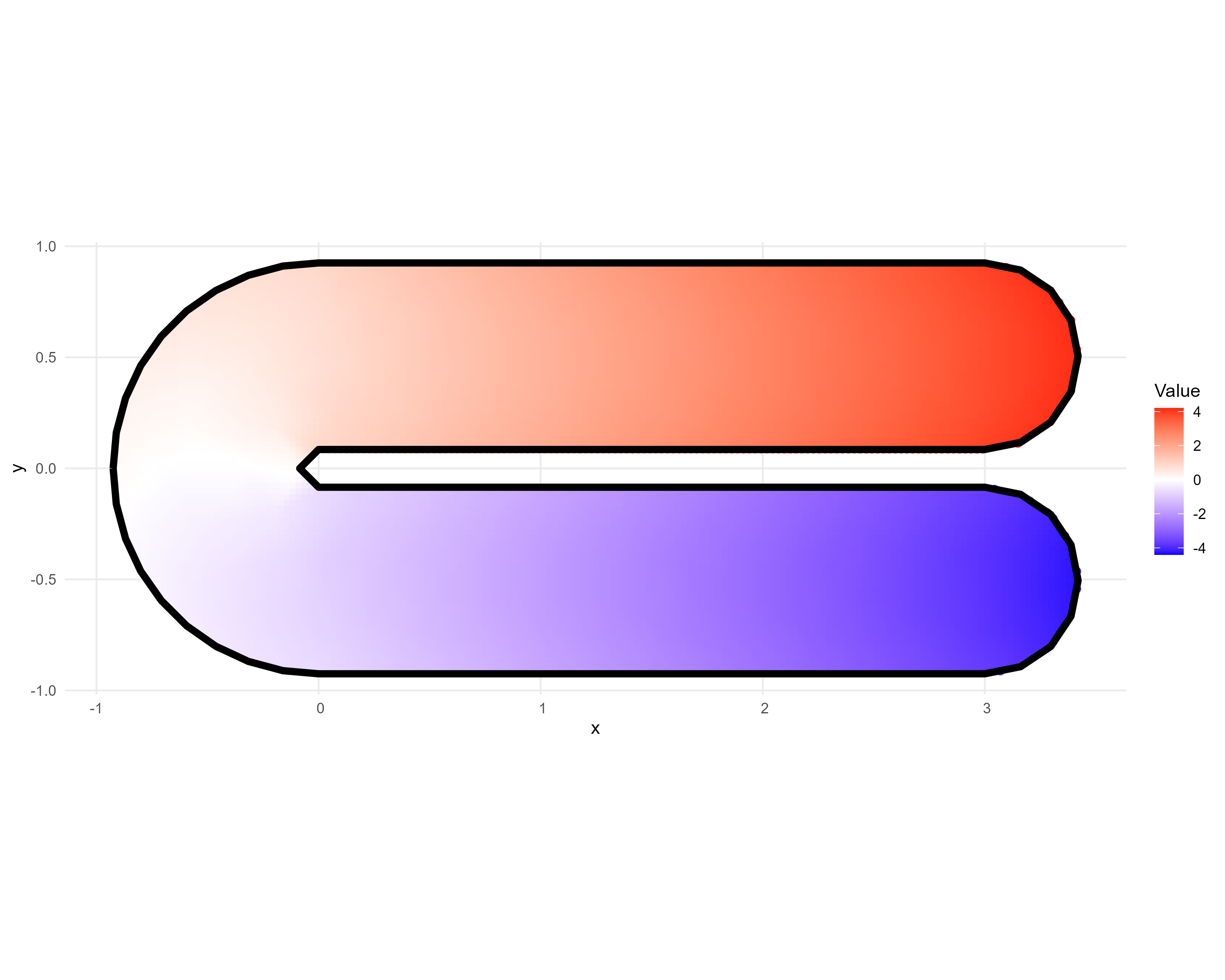}
        \par\vspace{0.3em}
         $(a)$ 
        \label{fig:true_spatial_effect}
    \end{minipage}
    \hfill
    \begin{minipage}[t]{0.48\textwidth}
        \centering
        \includegraphics[width=\textwidth]{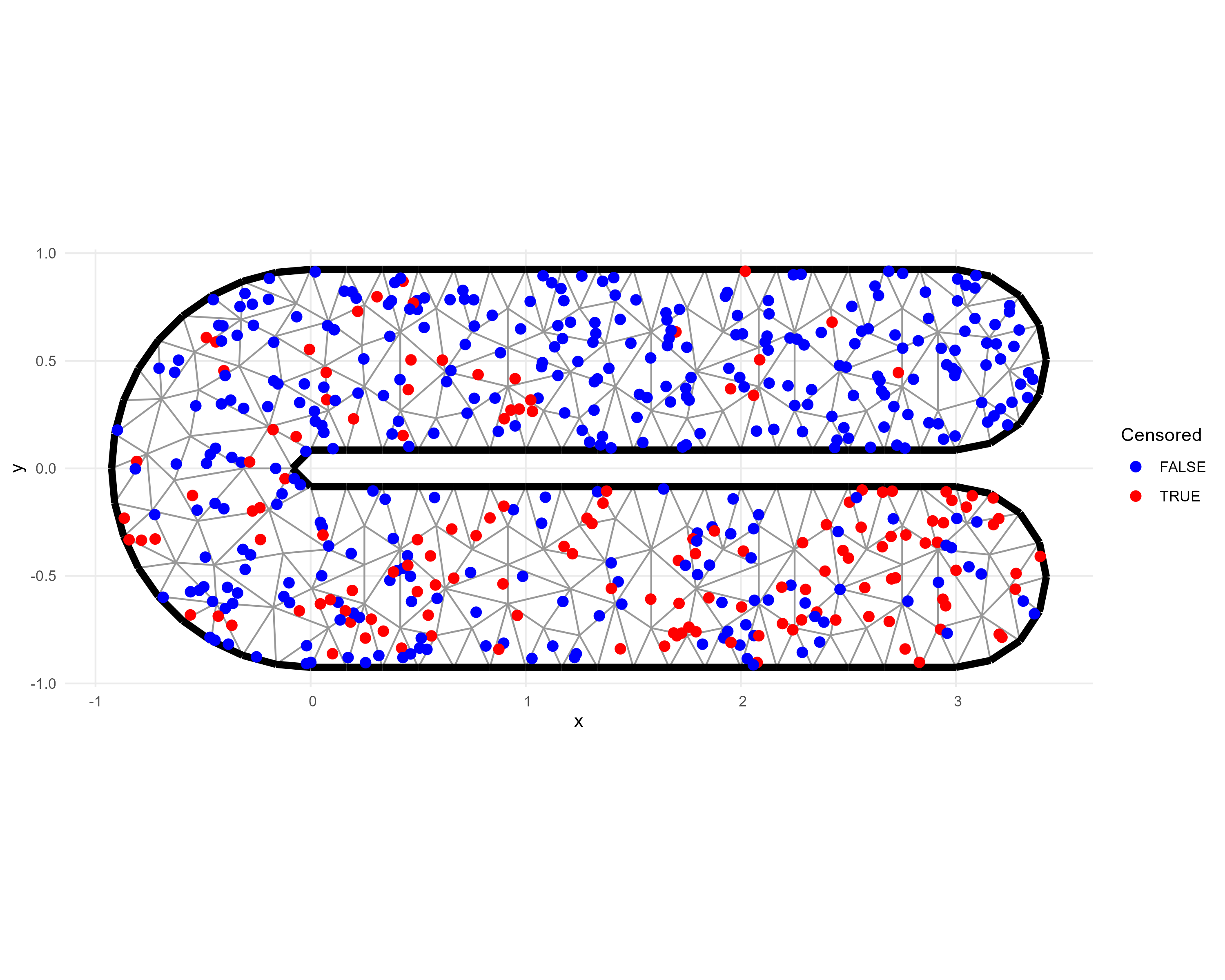}
        \par\vspace{0.3em}
         $(b)$ 
        \label{fig:sampled_points}
    \end{minipage}

    \vspace{1em}

    \begin{minipage}[t]{0.48\textwidth}
        \centering
        \includegraphics[width=\textwidth]{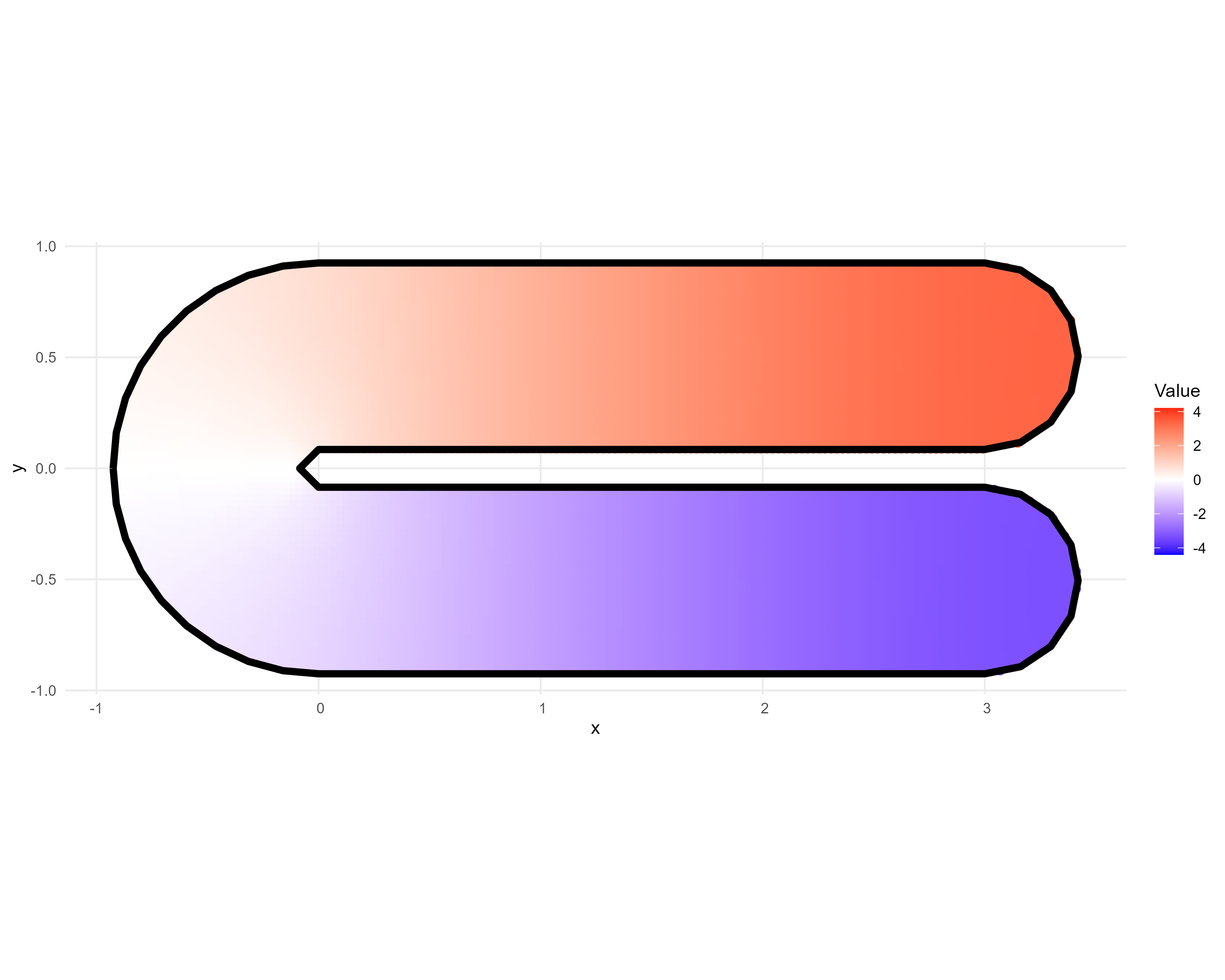}
        \par\vspace{0.3em}
         $(c)$
        \label{fig:approx_spatial_effect}
    \end{minipage}
    \hfill
    \begin{minipage}[t]{0.48\textwidth}
        \centering
        \includegraphics[width=\textwidth]{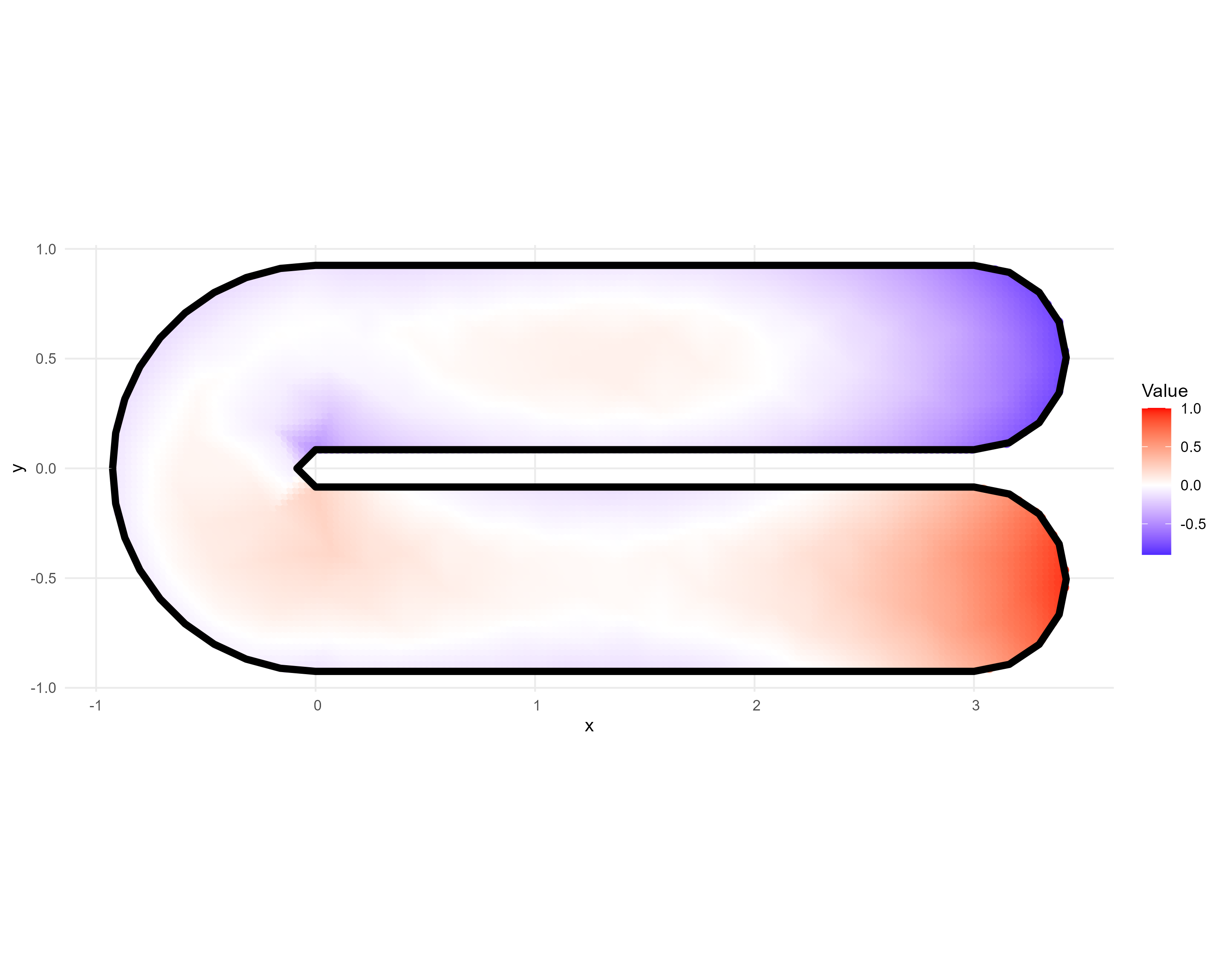}
        \par\vspace{0.3em}
         $(d)$ 
        \label{fig:error_effect}
    \end{minipage}

    \caption{ Spatial simulation results: $(a)$ true spatial effect \(h_0\) over the horseshoe-shaped domain \(\Omega\); $(b)$ sampling design and finite-element mesh for \(n = 500\) and 30\% censoring; $(c)$ estimated spatial effect \(\hat h\), averaged over \(N = 250\) replications; $(d)$ average pointwise error surface, \(h_0 - \hat h\) for \(n = 500\) with 30\% censoring.}
    \label{fig:all_spatial_effects}
\end{figure}

To evaluate the performance of the proposed model, we conducted simulations on a horseshoe-shaped domain $\Omega$ as described in \cite{ramsay2002spline} and \cite{wood2008soap}. The true spatial effect is displayed in Figure~\ref{fig:all_spatial_effects} $(a)$. 
Note that $\int_{\Omega} h_0(\bm p)d\bm p \approx 0.$ We observed independent replicates of
\(
(Y, \delta, \bm X, \bm P) = \left( \min(T, C),  I(T \le C),  \bm X,  \bm P \right),
\)
where \(\bm X = (X_1, X_2)\) has \(X_1 \sim \mathcal{N}(0,1)\) and \(X_2 \sim \text{Bernoulli}(0.5)\), and \(\bm P\) is drawn uniformly from a grid over the domain with spacing 0.025. Conditional on \((\bm X, \bm P)\), the event time \(T\) follows an exponential distribution with rate parameter
\( \exp(-4 \log(10) + \bm X^\top \bm\beta_0 + h_0(\bm P)), \)
where \(\bm\beta_0 = (0.25, -1)^\top\), matching the model in~\eqref{eq:model}.  
The censoring time \(C\) is exponentially distributed with rate varying in  \( \{0.27, 2.00\}\), corresponding to approximately \(15\%\) and \(30\%\) censoring, respectively. The analysis is conducted using sample sizes of \( n = 500, 1000, \) and \( 2000 \).

We construct the mesh using the package of \cite{sangalli2021spatial} and refine the domain so that the maximum area $\eta$ of the mesh triangles is consistent with the asymptotic analysis, setting $\eta = 0.1 n^{-0.45}$. We employ first-order FEM basis functions to reduce computational complexity.

An illustration for the case \( n = 500 \) is provided in Figure~\ref{fig:all_spatial_effects} $(b)$. 
Blue dots represent uncensored observations, red dots indicate censored observations, and the gray background depicts the constructed mesh.

We performed \(N = 250\) Monte Carlo replications. The penalized partial likelihood in~\eqref{eq:penalized_likelihood} was maximized using the quasi-Newton algorithm described in Section~\ref{sec:implementation}. 

At each replication, the regularization parameter \(\lambda\) was selected via the CV-PLD procedure (Section~\ref{sec:selection}) over the grid $\Lambda_n = \{\lambda_j = |\Omega|\,n^{-0.55}e^{\ell_j},\ \ell_j = \log(0.05) + \tfrac{j-1}{9}(\log(50) - \log(0.05)),\ j = 1,\dots,10\}$, where $|\Omega|$ indicates the area of $\Omega$ and normalizes the Laplacian penalty so that the regularization strength is independent of the size of the spatial domain. The optimal value \(\hat{\lambda} \in \Lambda_n\) was determined using five-fold cross-validation. 

For the illustrative example, Figure~\ref{fig:all_spatial_effects} $(c)$ 
displays the average estimated spatial effect across the \(N = 250\) Monte Carlo replications for $n = 500$ and 30\% of censoring rate, while Figure~\ref{fig:all_spatial_effects} $(d)$
shows the corresponding pointwise mean estimation error. The method accurately recovers the spatial effect, though higher errors appear in the lower branch of the domain where censoring is more concentrated and near the boundary.

Table~\ref{tab:performance-metrics} reports the estimation performance of the proposed spatial PH model in comparison with the standard PH model and the GAM PH model. In the latter, spatial variation is accommodated by a two-dimensional thin plate regression spline basis, specified as $s(x,y)$ with basis dimension $k=100$.

Reported metrics include coefficient bias, mean squared error (MSE) for regression parameters, defined as 
\( \mathrm{MSE} = N^{-1}\sum_{i=1}^N \|\hat{\boldsymbol{\beta}}^{\,i} - \boldsymbol{\beta}_0\|^2,\) where $\hat{\boldsymbol{\beta}}^{\,i}$ is the estimate from the $i$-th Monte Carlo replication, $L^2$ norms of bias for the spatial effect estimates, and 95\% coverage probabilities for the components of $\boldsymbol{\beta}$ constructed based on the Hessian and normal approximation.

Overall, the simulation results emphasize three main insights. First, as the sample size increases and the censoring rate decreases, performance improves markedly: both regression coefficient MSE and $L^2$ bias of the spatial effect shrink, confirming consistency. Second, the spatial PH estimator consistently outperforms the GAM approach in terms of spatial effect recovery, with substantially smaller $L^2$ bias across all scenarios. Third, both GAM approach and the standard PH model exhibit persistent bias in $\bm \beta$ regardless of sample size or censoring, underscoring model misspecification. Lastly, the coverage proabbility are near to the theoretical values. Taken together, these findings highlight the robustness of the proposed spatial PH method and its clear advantages in capturing both covariate effects and spatial structure.

\begin{table}[h]
\begingroup  
\renewcommand{\arraystretch}{1.0}  
\setlength{\arrayrulewidth}{0.4pt}  
\centering
\setlength{\tabcolsep}{3.5pt}
\footnotesize
\begin{adjustbox}{max width=\textwidth, center}
\begin{tabular}{cc|rrrrll}
\cline{3-8}
\multicolumn{1}{l}{}                        & \multicolumn{1}{l|}{} & \multicolumn{6}{c|}{\textbf{Proposed Spatial PH}}                                                                                                                                                                                                                                         \\ \hline
\multicolumn{1}{|c|}{\textbf{Sample Size}}  & \textbf{Censor Level} & \multicolumn{1}{c|}{\textbf{Bias $\beta_1$}} & \multicolumn{1}{c|}{\textbf{Bias $\beta_2$}} & \multicolumn{1}{c|}{\textbf{MSE $\bm\beta$}} & \multicolumn{1}{c|}{\textbf{$L_2$ Bias $h_0$}} & \multicolumn{1}{c|}{\textbf{CP95 $\beta_1$}} & \multicolumn{1}{l|}{\textbf{CP95 $\beta_2$}} \\ \hline
\multicolumn{1}{|c|}{\multirow{2}{*}{500}}  & 15\%                  & \multicolumn{1}{r|}{-0.003}                  & \multicolumn{1}{r|}{0.017}                   & \multicolumn{1}{r|}{0.011}                   & \multicolumn{1}{r|}{0.251}                     & \multicolumn{1}{r|}{0.952}                   & \multicolumn{1}{r|}{0.944}                   \\ \cline{2-8} 
\multicolumn{1}{|c|}{}                      & 30\%                  & \multicolumn{1}{r|}{-0.003}                  & \multicolumn{1}{r|}{0.021}                   & \multicolumn{1}{r|}{0.014}                   & \multicolumn{1}{r|}{0.293}                     & \multicolumn{1}{r|}{0.948}                   & \multicolumn{1}{r|}{0.936}                   \\ \hline
\multicolumn{1}{|c|}{\multirow{2}{*}{1000}} & 15\%                  & \multicolumn{1}{r|}{-0.002}                  & \multicolumn{1}{r|}{0.019}                   & \multicolumn{1}{r|}{0.006}                   & \multicolumn{1}{r|}{0.135}                     & \multicolumn{1}{r|}{0.960}                   & \multicolumn{1}{r|}{0.936}                   \\ \cline{2-8} 
\multicolumn{1}{|c|}{}                      & 30\%                  & \multicolumn{1}{r|}{-0.002}                  & \multicolumn{1}{r|}{0.023}                   & \multicolumn{1}{r|}{0.007}                   & \multicolumn{1}{r|}{0.161}                     & \multicolumn{1}{r|}{0.960}                   & \multicolumn{1}{r|}{0.944}                   \\ \hline
\multicolumn{1}{|c|}{\multirow{2}{*}{2000}} & 15\%                  & \multicolumn{1}{r|}{-0.001}                  & \multicolumn{1}{r|}{0.001}                   & \multicolumn{1}{r|}{0.003}                   & \multicolumn{1}{r|}{0.084}                     & \multicolumn{1}{r|}{0.932}                   & \multicolumn{1}{r|}{0.952}                   \\ \cline{2-8} 
\multicolumn{1}{|c|}{}                      & 30\%                  & \multicolumn{1}{r|}{-0.001}                  & \multicolumn{1}{r|}{0.001}                   & \multicolumn{1}{r|}{0.003}                   & \multicolumn{1}{r|}{0.096}                     & \multicolumn{1}{r|}{0.936}                   & \multicolumn{1}{r|}{0.944}                   \\ \hline
\multicolumn{1}{l}{}                        & \multicolumn{1}{l|}{} & \multicolumn{4}{c|}{\textbf{GAM PH}}                                                                                                                                                        &                                              &                                              \\ \cline{1-6}
\multicolumn{1}{|c|}{\multirow{2}{*}{500}}  & 15\%                  & \multicolumn{1}{r|}{-0.049}                  & \multicolumn{1}{r|}{0.201}                   & \multicolumn{1}{r|}{0.054}                   & \multicolumn{1}{r|}{3.356}                     &                                              &                                              \\ \cline{2-6}
\multicolumn{1}{|c|}{}                      & 30\%                  & \multicolumn{1}{r|}{-0.098}                  & \multicolumn{1}{r|}{0.395}                   & \multicolumn{1}{r|}{0.178}                   & \multicolumn{1}{r|}{7.731}                     &                                              &                                              \\ \cline{1-6}
\multicolumn{1}{|c|}{\multirow{2}{*}{1000}} & 15\%                  & \multicolumn{1}{r|}{-0.049}                  & \multicolumn{1}{r|}{0.203}                   & \multicolumn{1}{r|}{0.049}                   & \multicolumn{1}{r|}{2.739}                     &                                              &                                              \\ \cline{2-6}
\multicolumn{1}{|c|}{}                      & 30\%                  & \multicolumn{1}{r|}{-0.097}                  & \multicolumn{1}{r|}{0.393}                   & \multicolumn{1}{r|}{0.170}                   & \multicolumn{1}{r|}{6.948}                     &                                              &                                              \\ \cline{1-6}
\multicolumn{1}{|c|}{\multirow{2}{*}{2000}} & 15\%                  & \multicolumn{1}{r|}{-0.050}                  & \multicolumn{1}{r|}{0.198}                   & \multicolumn{1}{r|}{0.045}                   & \multicolumn{1}{r|}{2.473}                     &                                              &                                              \\ \cline{2-6}
\multicolumn{1}{|c|}{}                      & 30\%                  & \multicolumn{1}{r|}{-0.096}                  & \multicolumn{1}{r|}{0.384}                   & \multicolumn{1}{r|}{0.160}                   & \multicolumn{1}{r|}{6.494}                     &                                              &                                              \\ \cline{1-6}
\multicolumn{1}{l}{}                        & \multicolumn{1}{l|}{} & \multicolumn{3}{c|}{\textbf{Standard PH}}                                                                                                  & \multicolumn{1}{l}{}                           &                                              &                                              \\ \cline{1-5}
\multicolumn{1}{|c|}{\multirow{2}{*}{500}}  & 15\%                  & \multicolumn{1}{r|}{-0.153}                  & \multicolumn{1}{r|}{0.614}                   & \multicolumn{1}{r|}{0.411}                   & \multicolumn{1}{l}{}                           &                                              &                                              \\ \cline{2-5}
\multicolumn{1}{|c|}{}                      & 30\%                  & \multicolumn{1}{r|}{-0.151}                  & \multicolumn{1}{r|}{0.610}                   & \multicolumn{1}{r|}{0.407}                   & \multicolumn{1}{l}{}                           &                                              &                                              \\ \cline{1-5}
\multicolumn{1}{|c|}{\multirow{2}{*}{1000}} & 15\%                  & \multicolumn{1}{r|}{-0.153}                  & \multicolumn{1}{r|}{0.600}                   & \multicolumn{1}{r|}{0.389}                   & \multicolumn{1}{l}{}                           &                                              &                                              \\ \cline{2-5}
\multicolumn{1}{|c|}{}                      & 30\%                  & \multicolumn{1}{r|}{-0.152}                  & \multicolumn{1}{r|}{0.593}                   & \multicolumn{1}{r|}{0.382}                   & \multicolumn{1}{l}{}                           &                                              &                                              \\ \cline{1-5}
\multicolumn{1}{|c|}{\multirow{2}{*}{2000}} & 15\%                  & \multicolumn{1}{r|}{-0.153}                  & \multicolumn{1}{r|}{0.611}                   & \multicolumn{1}{r|}{0.400}                   & \multicolumn{1}{l}{}                           &                                              &                                              \\ \cline{2-5}
\multicolumn{1}{|c|}{}                      & 30\%                  & \multicolumn{1}{r|}{-0.152}                  & \multicolumn{1}{r|}{0.606}                   & \multicolumn{1}{r|}{0.394}                   & \multicolumn{1}{l}{}                           &                                              &                                              \\ \cline{1-5}
\end{tabular}
\end{adjustbox}
\endgroup  
\caption{Estimation performance of the proposed spatial PH model compared with the standard PH model and the GAM PH model across different sample size ($n$) and censoring levels. Reported metrics include coefficient bias, mean squared error (MSE) of the regression parameters, $L_2$ norm of bias for the spatial effect estimates (only applicable to models with spatial components), and empirical 95\% coverage probabilities (CP95) for the regression coefficients in the proposed model.}
\label{tab:performance-metrics}
\end{table}

\FloatBarrier
\section{Empirical Application}\label{sec:empirical} In this section, we present two empirical studies that illustrate the practical use of the proposed methodology. The first examines emergency medical service (EMS) response times for ambulance units in the city of San Francisco, while the second focuses on crowdsourced seismic data collected during an earthquake in the Campi Flegrei area, Italy. Together, these applications highlight the flexibility of the model across very different domains: urban public safety and real-time earthquake monitoring. Both cases demonstrate the model's ability to handle complex spatial domains—ranging from the intricate coastline of a peninsula to the volcanic topography of a caldera—while adjusting for high-dimensional temporal and environmental covariates.

\subsection{San Francisco EMS Ambulance response-time}
The proposed methodology is illustrated using San Francisco Division of Emergency Medical Services (EMS) dispatch data of February 2025, comprising 7,890 different interventions. Each intervention records a sequence of chronological milestones, including the time the call was received, the unit dispatch time, and the start of the unit’s response. The event of interest is the duration until the first unit’s arrival on scene, which is subject to right-censoring originating from two distinct sources. First, operational censoring occurs when a unit is diverted to a higher-priority incident or cancelled before reaching the scene. In such instances, the arrival timestamp is missing, and the censoring time is defined as the duration from the initial call to the last recorded operational activity. Second, administrative censoring was applied to limit the influence of extreme outliers by capping all response times at 60 minutes. For reference, the 99th percentile of response time is approximately 37 minutes. Any intervention exceeding 60 minutes was therefore treated as right-censored at that threshold.

The model is adjusted for several environmental and temporal covariates that might influence medical unit mobility. Weather data were sourced from the Open-Meteo archive, and is relative to the average coordinates of the observed locations (37.77$^\circ$ N, 122.42$^\circ$ W).  Weather data includes temperature (\texttt{temp}) measured in degrees Celsius, hourly precipitation (\texttt{precip}) in millimeters, and wind speed (\texttt{wind}) in kilometers per hour. Daily periodicity is captured via harmonic terms $s_k, c_k$ for $k=1,2$, representing 24-hour and 12-hour cycles derived from the decimal hour of the call. We also include a binary indicator for high-priority emergency calls (\texttt{is\_emergency}), and a holiday indicator (\texttt{is\_holiday}) based on the New York Stock Exchange (NYSE) calendar. The latter serves as a proxy for urban traffic dynamics, as NYSE holidays align with the closure of San Francisco’s financial district and school systems, significantly reducing road congestion.

\begin{figure}[htp]
    \centering

   \begin{minipage}[t]{0.42\textwidth}
        \centering
        \includegraphics[width=\textwidth,keepaspectratio]{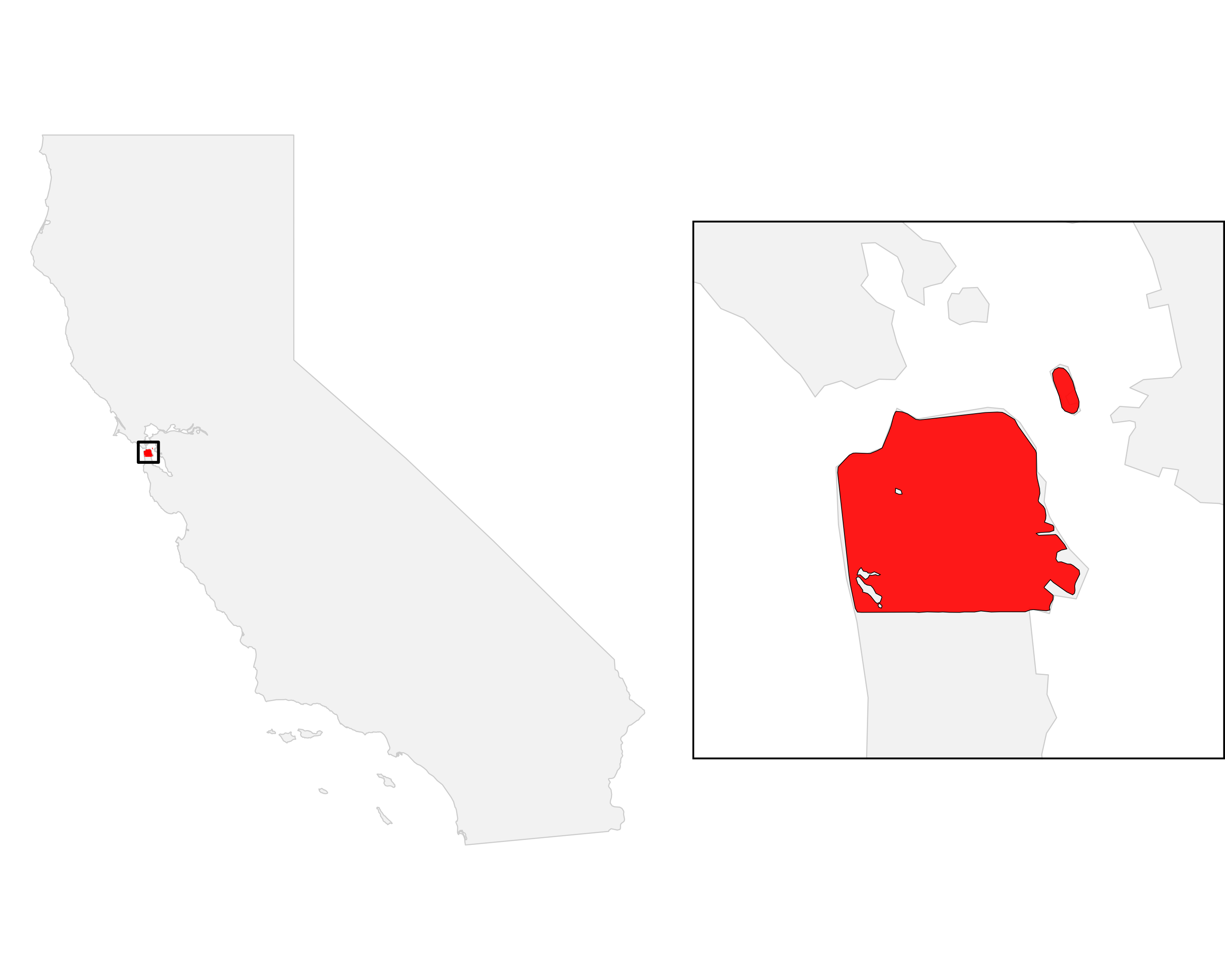}
        \par\vspace{0.35em}
        \small $(a)$ 
        \label{fig:sf:inset}
    \end{minipage}
    \hfill
    \begin{minipage}[t]{0.52\textwidth}
        \centering
        \includegraphics[width=\textwidth,keepaspectratio]{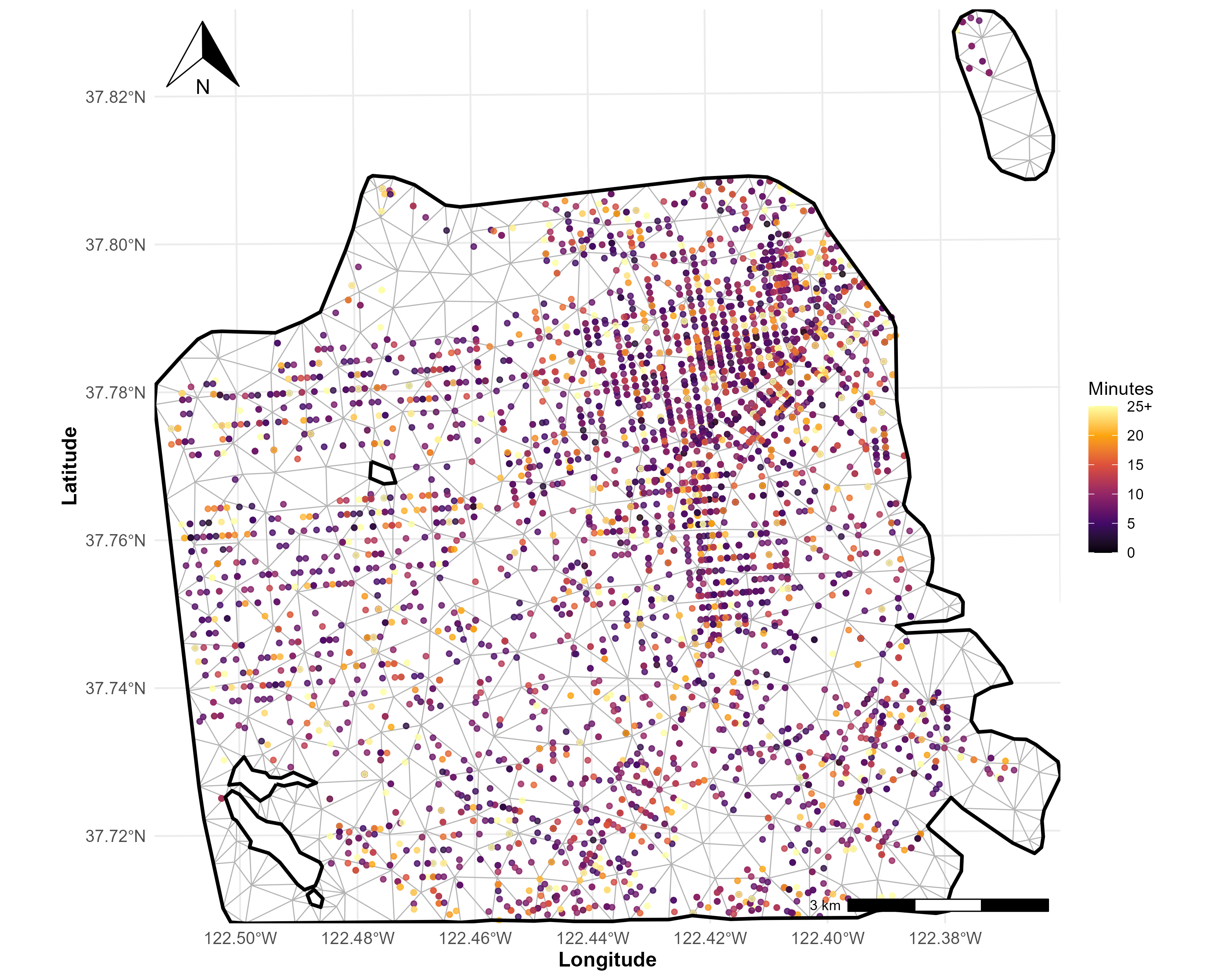}
        \par\vspace{0.35em}
        \small $(b)$ 
        \label{fig:sf:mesh}
    \end{minipage}

    \caption{Spatial domain and data for San Francisco in February 2025: $(a)$ location of the San Francisco study area within California; $(b)$ finite element mesh and EMS response-time observations, with points indicating incident locations colored according to observed time.}
    \label{fig:sf_combined}
\end{figure}

In order to accommodate San Francisco’s complex coastal geometry and internal water bodies, we define the spatial domain of interest as the city’s geographic boundaries where major water features, such as Lake Merced, are excluded, while Treasure Island is retained. The resulting domain is shown in Figure \ref{fig:sf_combined} (a), displayed in the context of the state of California.  This domain is then discretized using a two-dimensional finite element mesh generated via constrained Delaunay refinement. The resulting triangulation consists of 468 nodes. Figure~\ref{fig:sf_combined} (b) displays the resulting mesh along with the spatial distribution of EMS incidents, where point colours indicate observed response times in minutes.

We apply the proposed method, selecting the smoothing parameter \( \lambda \) as in the simulation study. The estimated regression coefficients \( \hat{\beta} \) are reported in Table~\ref{tab:sf_empirical_betas}. The reported numerical values correspond to the estimated coefficients, while the relative \( p \)-values are computed using standard errors obtained from the inverse Hessian of the penalized log-likelihood.

We observe that the indicator variable \texttt{is\_emergency} is the most dominant predictor, with an estimated coefficient of 1.964 \(( p < 0.001 )\). This implies a substantially higher arrival hazard for emergency-status calls. The coefficient for \texttt{precip} is 0.035 and is marginally significant \(( p = 0.064 )\), suggesting a modest increase in the hazard under precipitation. Furthermore, the 12-hour harmonic component \( s_2 \) is statistically significant \(( p = 0.030 )\), indicating a semi-diurnal rhythm in EMS efficiency, possibly associated with peak demand periods or shift changes. The estimated effects of \texttt{wind}, \texttt{temp}, and \texttt{is\_holiday} are instead not statistically different from zero at the 10\% significance level.

\begin{table}[H]

	\centering
    \small
	\begin{tabular}{l|rrr|}
		\cline{2-4}
		& \multicolumn{3}{c|}{Proposed Spatial PH} \\ \hline
		\multicolumn{1}{|l|}{\textbf{Covariate}} 
		& \multicolumn{1}{r|}{\textbf{Estimate}} 
		& \multicolumn{1}{r|}{\textbf{Std. Error}} 
		& \textbf{p-value} \\ \hline
		\multicolumn{1}{|l|}{is\_mergency} & \multicolumn{1}{r|}{$1.964$} & \multicolumn{1}{r|}{$0.028$} & $<0.001$ \\ \hline
		\multicolumn{1}{|l|}{temp}      & \multicolumn{1}{r|}{$-0.002$} & \multicolumn{1}{r|}{$0.005$} & $0.685$ \\ \hline
		\multicolumn{1}{|l|}{precip}    & \multicolumn{1}{r|}{$0.035$}  & \multicolumn{1}{r|}{$0.019$} & $0.064$ \\ \hline
		\multicolumn{1}{|l|}{wind}      & \multicolumn{1}{r|}{$-0.002$} & \multicolumn{1}{r|}{$0.002$} & $0.300$ \\ \hline
		\multicolumn{1}{|l|}{is\_holiday}& \multicolumn{1}{r|}{$0.038$}  & \multicolumn{1}{r|}{$0.026$} & $0.150$ \\ \hline
		\multicolumn{1}{|l|}{$s_1$ (24h)} & \multicolumn{1}{r|}{$-0.007$} & \multicolumn{1}{r|}{$0.018$} & $0.680$ \\ \hline
		\multicolumn{1}{|l|}{$c_1$ (24h)} & \multicolumn{1}{r|}{$-0.022$} & \multicolumn{1}{r|}{$0.021$} & $0.305$ \\ \hline
		\multicolumn{1}{|l|}{$s_2$ (12h)} & \multicolumn{1}{r|}{$-0.037$} & \multicolumn{1}{r|}{$0.017$} & $0.030$ \\ \hline
		\multicolumn{1}{|l|}{$c_2$ (12h)} & \multicolumn{1}{r|}{$0.004$}  & \multicolumn{1}{r|}{$0.017$} & $0.814$ \\ \hline
	\end{tabular}
\caption{Estimated covariate effects from the spatial proportional hazards model for San Francisco EMS ambulance response times (February 2025). Estimates are log-hazard ratios; $s_k$ and $c_k$ denote sine and cosine terms capturing 24-hour ($k=1$) and 12-hour ($k=2$) temporal cycles.}
\label{tab:sf_empirical_betas}
\end{table}

\begin{figure}[htp]
    \centering
    \includegraphics[width=0.7\textwidth]{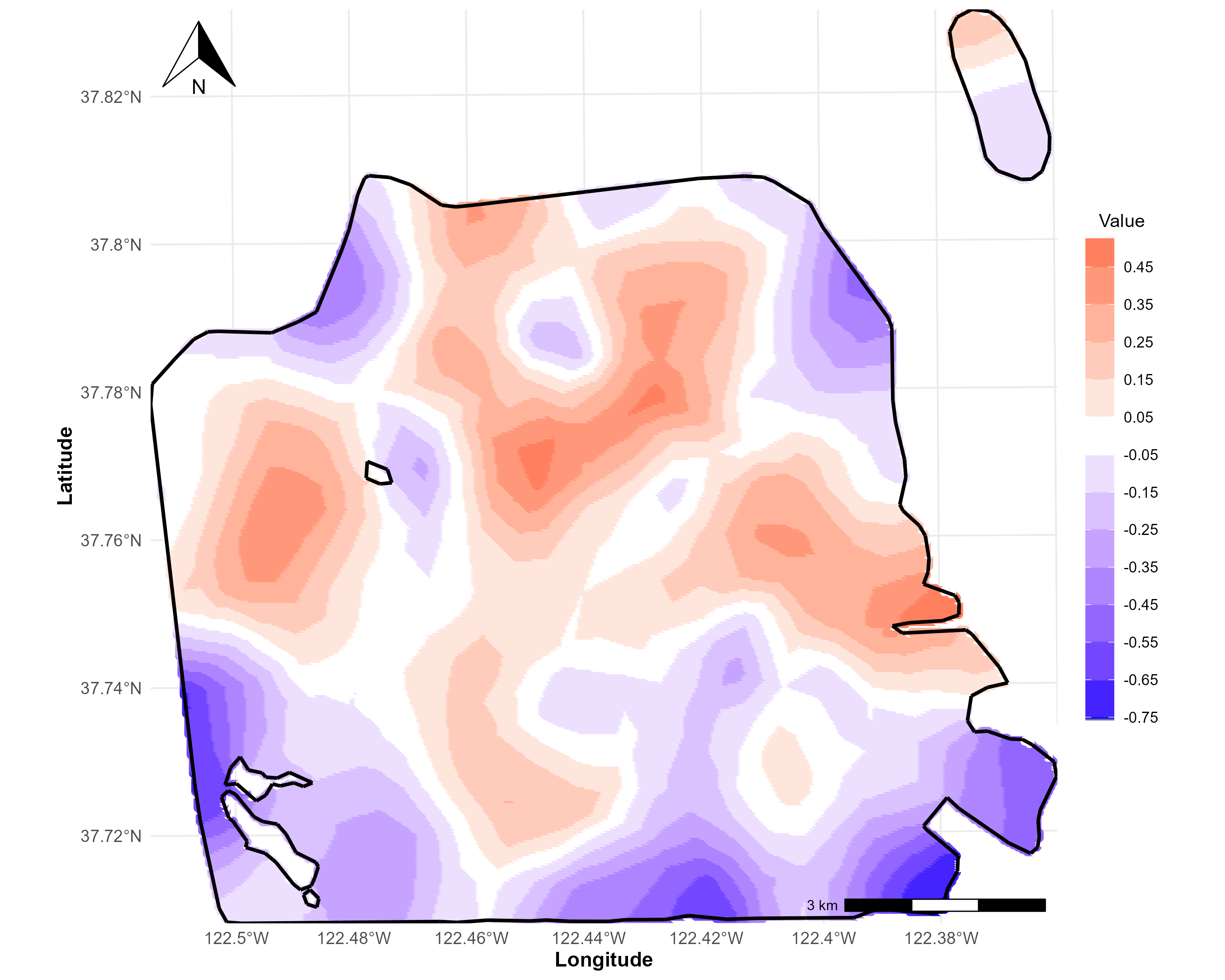}
    \caption{Estimated spatial effect $\hat{h}(\bm{p})$, where higher values correspond to faster expected emergency unit arrivals.}
    \label{fig:sf_spatial_effect}
\end{figure}

Beyond these global trends, the model uncovers geographic disparities. Figure~\ref{fig:sf_spatial_effect} illustrates the estimated spatial function $\hat{h}(\bm{p})$, where lower values indicate weaker response efficiency. These variations persist after adjusting for covariates, indicating that structural characteristics of the urban environment influence expected arrival times. Higher arrival hazards (faster expected responses) are concentrated in central and north-eastern areas, whereas south-western and south-eastern neighbourhoods exhibit lower hazards. Treasure Island displays a clear north–south gradient, with relatively higher hazards in the north and lower hazards in the south. These disparities likely reflect differences in road-network configuration, station proximity, topographical constraints, and localised access conditions. The model thus provides a representation
of EMS risk, offering insights into the geographic and temporal  distribution of EMS arrival performance across the urban landscape.

\FloatBarrier
\subsection{Crowdsourced seismic data analysis}\label{sec:earthquake}  
Earthquake Network (EQN) is a citizen science initiative for real time earthquake monitoring using crowdsourced smartphones \citep{finazzi2016}. When a smartphone detects shaking, it sends a signal (called trigger) to a central server. Based on all the triggers received, the server decides whether an earthquake is occurring. If an earthquake is detected, an alert containing a warning about the incoming seismic wave is sent to the population.
Earthquake detection also represents a censoring event, meaning any subsequent triggers are ignored.

\begin{figure}[htp]
    \centering

    \begin{minipage}[t]{0.48\textwidth}
        \centering
        \includegraphics[width=\textwidth,keepaspectratio]{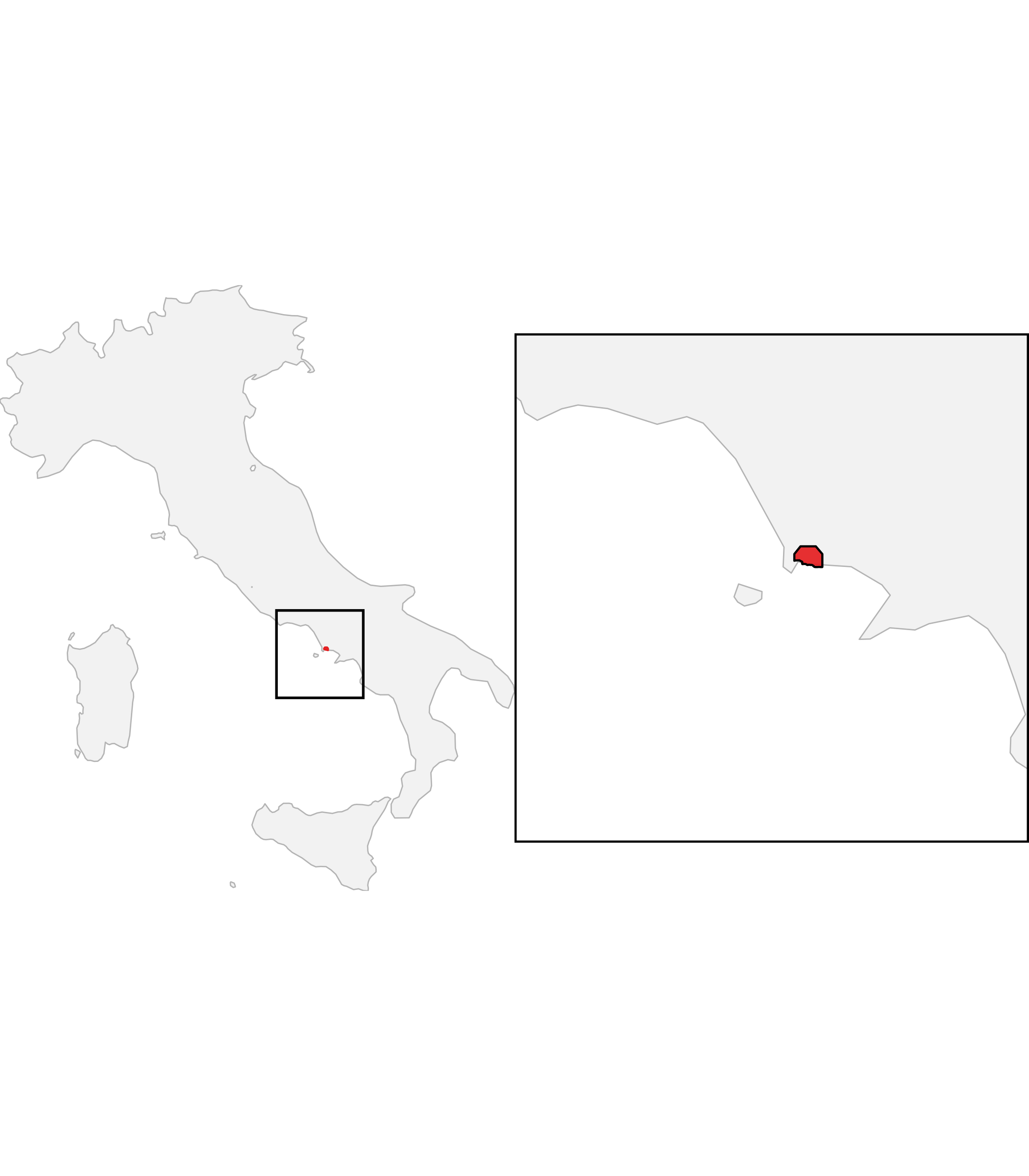}
        \par\vspace{0.35em}
        \small $(a)$
        \label{fig:data:inset}
    \end{minipage}
    \hfill
    \begin{minipage}[t]{0.48\textwidth}
        \centering
        \includegraphics[width=\textwidth,keepaspectratio]{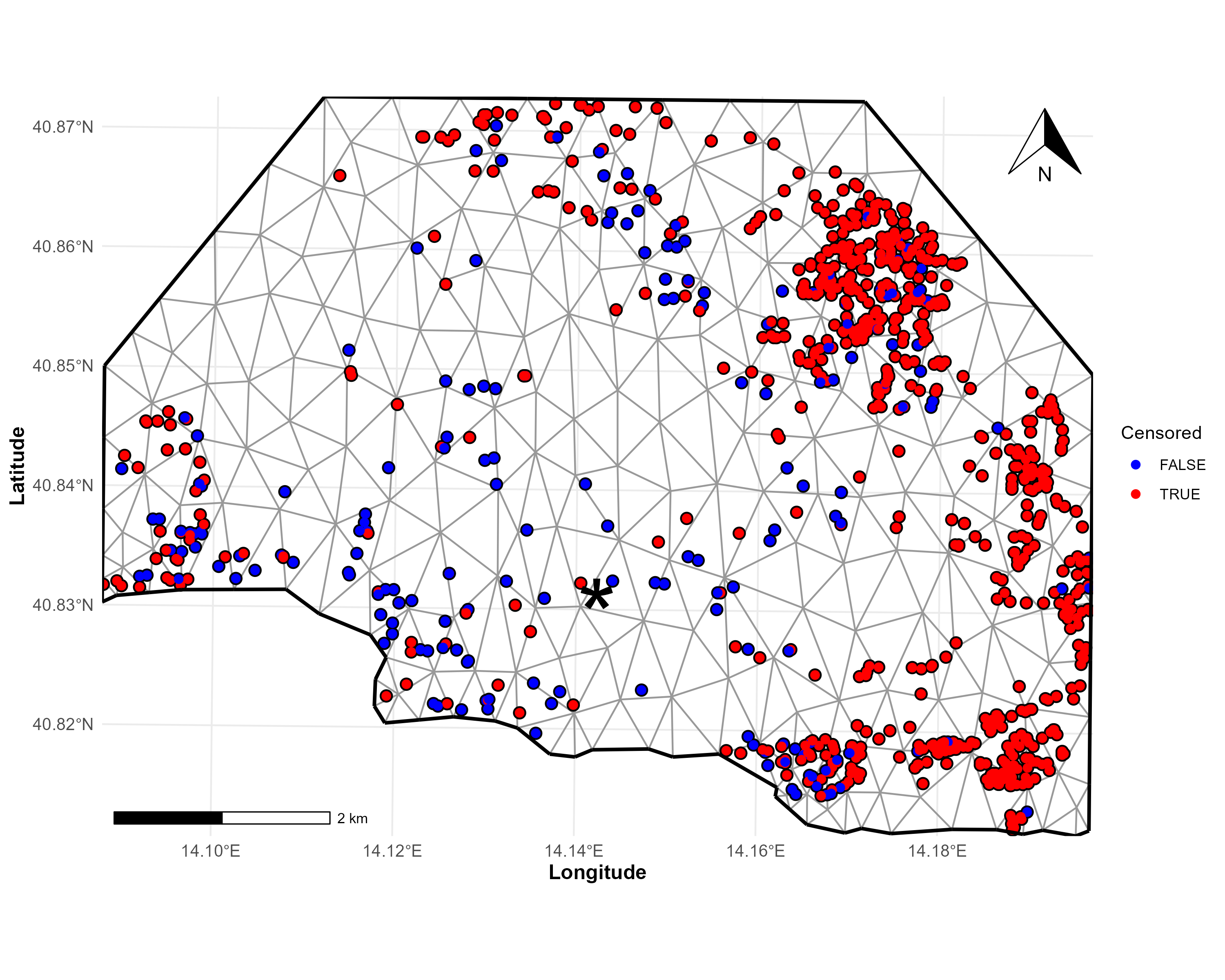}
        \par\vspace{0.35em}
        \small $(b)$
        \label{fig:data:mesh}
    \end{minipage}

    \vspace{1.5em} 

    \begin{minipage}[t]{0.55\textwidth} 
        \centering
        \includegraphics[width=\textwidth,keepaspectratio]{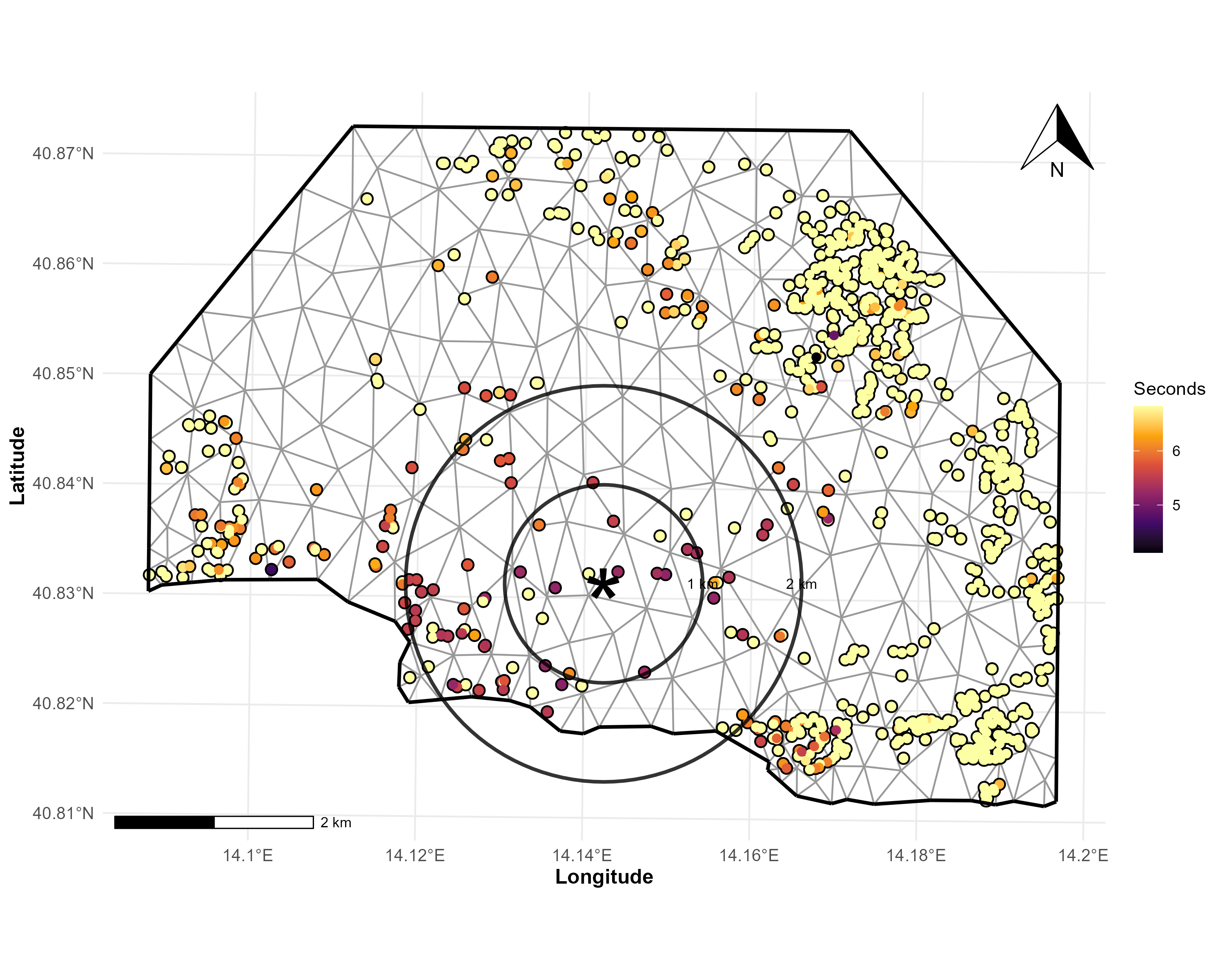}
        \par\vspace{0.35em}
        \small $(c)$
        \label{fig:data:time}
    \end{minipage}

    \caption{Saptial domain and data for Campi Flegrei earthquake data (18 Feb 2025): $(a)$ location of the study area within Italy, $(b)$ triangular mesh with censored (red) and uncensored (blue) observations, and $(c)$ observed smartphone triggering times with distance rings from the epicenter (star) colored according to observed time.}
    \label{fig:campi_flegrei_data}
\end{figure}

Here, we examine the EQN triggering times collected during the magnitude 3.1 earthquake occurred on 18 February 2025 at 02:22:19 UTC. The epicentre was located in the Campi Flegrei area (Italy), as shown in Figure \ref{fig:campi_flegrei_data}(a) at a depth of 2 km. 

We claim that triggering times carry information about site amplification, i.e. the amplification of seismic waves due to local geology. For a given distance from the epicentre, we expect smartphones to detect seismic waves slightly earlier where amplification is high, and slightly later where amplification is low (de-amplification). Additionally, we expect the number of censored smartphones to be higher in de-amplification areas because the smartphone may not detect the seismic wave at all.

For each smartphone \(i\) we observe \(Y_i\), which may be equal to the triggering time or to the time of the censoring event (i.e., the earthquake detection by EQN). The only model covariate is the hypocentral distance. Letting \((\text{lat}_i,\text{lon}_i)\) denote the smartphone coordinates, we compute the central angle \(\Delta\sigma_i\) using the haversine formula and define the 3D depth-adjusted distance
\[
\text{dist}_i = \sqrt{\,(\text{depth})^2 + \big(2R\sin(\Delta\sigma_i/2)\big)^2\,},
\]
where \(R=6371\,\text{km}\) is the Earth’s radius and depth \(=2\,\text{km}\). To ensure a fair proportion of uncensored values within the study area, we restrict the data to smartphones located within \(5\,\text{km}\) from the epicentre. This resulted in a sample of \(976\) triggering times, approximately \(80\%\) of which are right-censored.

As before, we construct a two-dimensional finite element mesh that conforms to the irregular boundary of the coastline of Campi Flegrei and the distance limit from the epicentre. The mesh is obtained by refining the boundary geometry and consists of $229$ vertices connected through triangular elements. The resulting mesh, along with smartphone locations colored by censoring status, is shown in Figure~\ref{fig:campi_flegrei_data} (b), while the observed triggering times are illustrated in Figure~\ref{fig:campi_flegrei_data} (c).

\begin{figure}[htp]
    \centering

    \begin{minipage}[t]{0.48\textwidth}
        \centering
        \includegraphics[width=\textwidth,keepaspectratio]{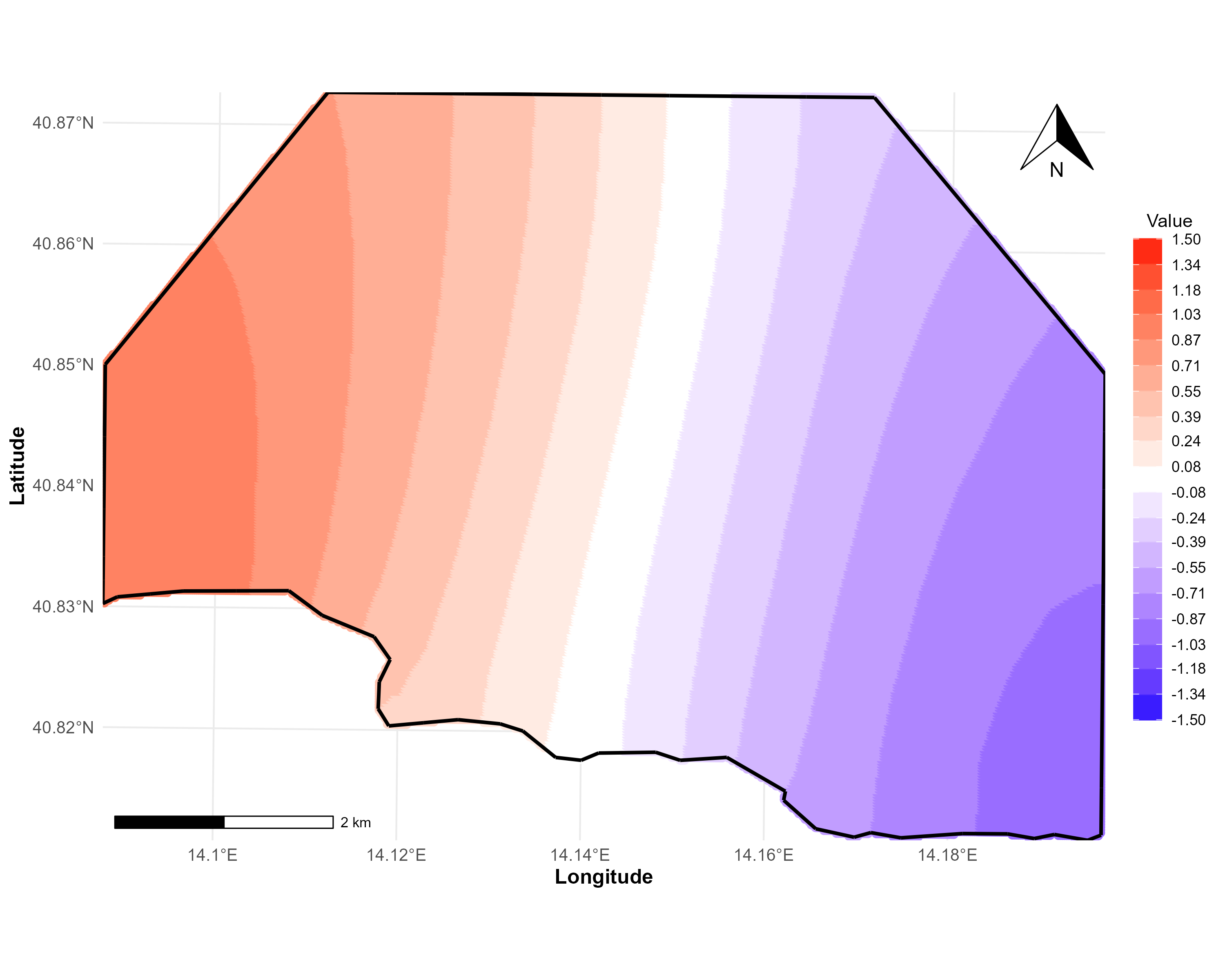}
        \par\vspace{0.35em}
        \small $(a)$
        \label{fig:results:spatial}
    \end{minipage}
    \hfill
    \begin{minipage}[t]{0.48\textwidth}
        \centering
        \includegraphics[width=\textwidth,keepaspectratio]{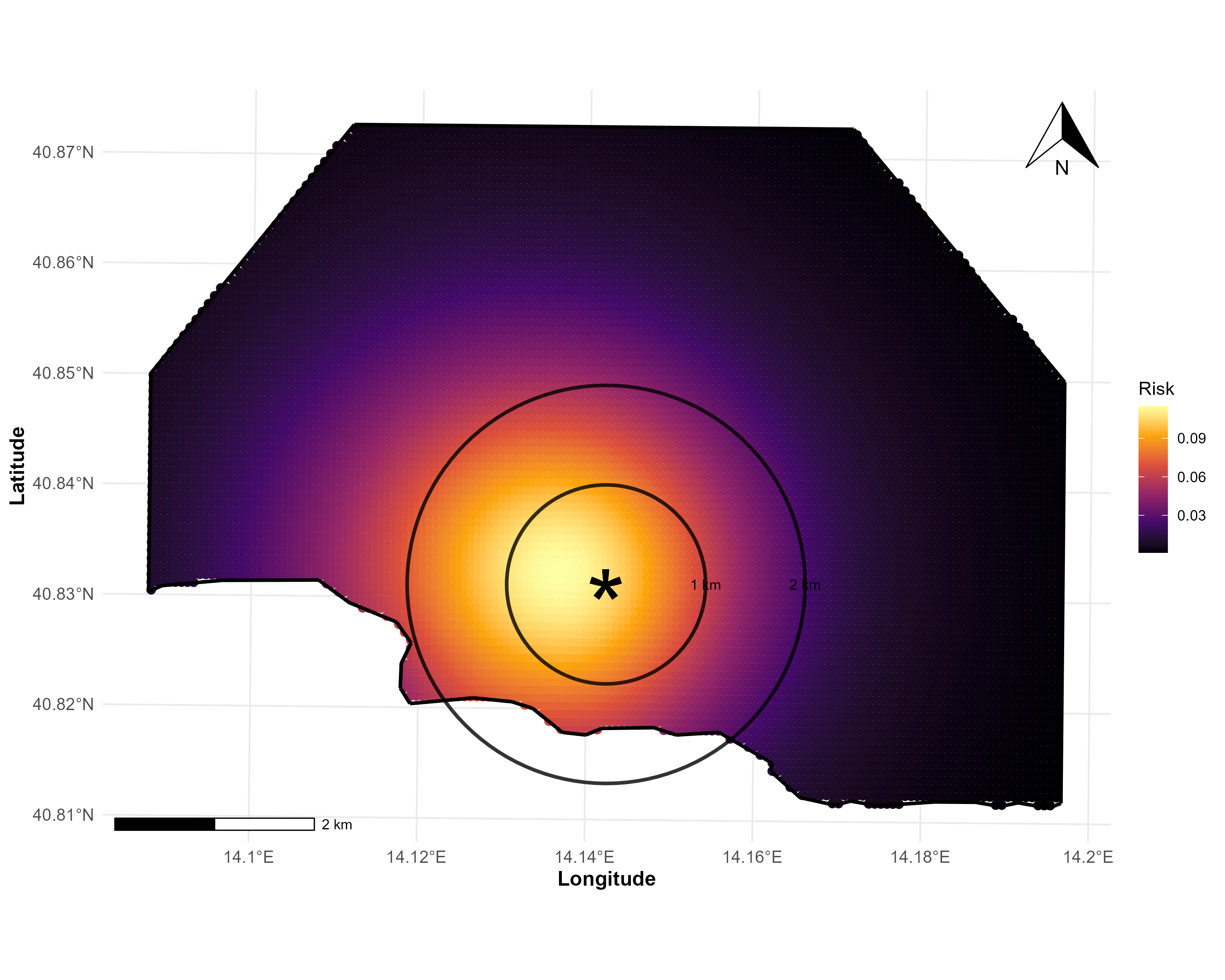}
        \par\vspace{0.35em}
        \small $(b)$ 
        \label{fig:results:risk}
    \end{minipage}

    \caption{Model results for the Campi Flegrei earthquake: $(a)$  estimated spatial effect \(\hat h(\bm p)\), where higher values indicate a greater hazard (faster triggering);  $(b)$ estimated relative hazard \(\exp(\bm x^\top \hat{\bm\beta} + \hat h(\bm p))\) at 12:00 PM, where warmer colours denote higher values.}
    \label{fig:campi_flegrei_results}
\end{figure}

The estimation procedure and the the smoothing parameter followed the same approach as in Section~\ref{sec:simulations}.
The estimated hypocentral distance effect is equal to $-1.124$, with standard error $0.102$ and p-value $<0.01$. This confirms that, as expected, smartphones located further from the epicentre have a lower hazard of triggering, that is, they tend to trigger later or not at all. On the other hand, Figure~\ref{fig:campi_flegrei_results} $(a)$ 
displays the estimated spatial effect after accounting for the hypocentral distance. The model uncovers a clear east–west gradient across the Campi Flegrei area, which is consistent with the findings of \cite{finazzi2025} who obtained the amplification map for the Campi Flegrei area from the analysis of the smartphone shaking intensity rather than triggering time. 
Finally, Figure~\ref{fig:campi_flegrei_results} $(b)$ 
illustrates the estimated spatial variation of the relative hazard for the specific earthquake. The highest hazard values are obviously concentrated around the epicentre. However, due to the spatial effect considered in the model (which describes site amplification), the gradient of the hazard is not radial with respect to the epicentre.

\FloatBarrier
\section{Conclusions}\label{sec:conclusion} 

We introduced a non-parametric proportional hazards model that incorporates a smooth spatial effect within the classical partial likelihood framework. By combining a Laplacian-based differential penalty with finite element sieve approximations, our approach can handle complex and irregular spatial domains while maintaining a concave optimisation problem and producing interpretable regression coefficients.

From a theoretical perspective, we proved that the model is identifiable, that the sieve estimator is consistent, and that the parametric component is asymptotically normal. These results ensure that standard inferential procedures for the regression coefficients remain valid despite the presence of an infinite-dimensional spatial effect. Empirically, simulation studies demonstrate that the proposed method accurately reconstructs latent spatial variation, reducing bias and mean squared error relative to standard proportional hazards models and yielding high-resolution, interpretable spatial risk surfaces. Two empirical applications further illustrate how explicitly modelling spatial heterogeneity can reveal meaningful structures in the data-generating process that would otherwise remain unobserved.

The proposed methodology, developed for spatial survival analysis, naturally extends to other spatial data structures such as areal and network data, and more broadly to settings requiring non-parametric effects within a proportional hazards framework. The finite element formulation provides a flexible and principled way to represent smooth effects over complex domains, making the approach applicable to a wide range of problems beyond those considered here.

There are several avenues for future research. One important area for further research concerns the goodness-of-fit of the proportional hazards assumption in the presence of a non-parametric spatial effect. Classical diagnostic tools, such as Schoenfeld residuals, are derived under the assumption of purely parametric covariate effects and therefore require careful adaptation in this setting. The development of residual- or score-based diagnostics that explicitly account for the estimated spatial component would enable practitioners to evaluate time-varying effects and identify deviations from proportionality following model fitting. Other avenues to explore include formally quantifying the uncertainty of the spatial effect, selecting the penalty structure based on the data, and extending spatio-temporal survival models to include latent effects that evolve over time.

\FloatBarrier





\bibliographystyle{unsrtnat}
\bibliography{sample}

\appendix
	\section*{Appendix}
The appendix is organized as follows. The Supplementary Material is organized as follows. Section~\ref{sec:areal} explains how to adapt the model in case of areal data. Section~\ref{appendix:identification} provides the proof of Theorem~\ref{theo:identification} on model identifiability.  
Section~\ref{appendix:consistency} contains the proof of Theorem~\ref{theo:consistency} establishing consistency of the estimator.  
Section~\ref{appendix:normality} gives the proof of Theorem~\ref{thm:beta-CLT-rewrite}.  
Finally, Section~\ref{appendix:lemmas} collects auxiliary lemmas on functional space compactness, continuity, differentiability, and uniqueness results used throughout the proofs.  

\section{Areal Data}\label{sec:areal}
The proposed methodology has so far been developed for geostatistical survival data, where observations are associated with point-referenced spatial locations $\bm{p}_1, \dots, \bm{p}_n \in \Omega$. In many applications, however, survival outcomes are available only for areal units, corresponding to disjoint spatial subdomains of $\Omega$. Let $\{D_1, \dots, D_n\}$ denote such a collection of regions. For instance, in epidemiological or environmental studies, data may be aggregated at the level of administrative districts, census tracts, or neighborhoods. In this setting, the spatial effect cannot be evaluated at a point, but must instead be represented through an areal functional of the latent effect. A natural choice is the mean of $h$ over region $D_i$, leading to the model \[ \lambda(t \mid \bm x, D_i) \;=\; \lambda_0(t) \exp\!\left( \bm{x}^\top \bm{\beta}_0 \;+\; \frac{1}{|D_i|}\int_{D_i} h(\bm p)\,d\bm p \right), \] where $|D_i|$ denotes the area of region $D_i$. Estimation of $(\bm\beta,h)$ proceeds by minimizing a penalized criterion analogous to the one introduced in \eqref{eq:penalized_likelihood}, but with integrals over areal units in place of pointwise evaluations. The finite element framework naturally accommodates this extension: nodal evaluations of $h$ are replaced by weighted averages over the mesh elements contained in each $D_i$. In practice, this requires the construction of an incidence matrix that maps finite element basis functions to the regions $\{D_1,\dots,D_n\}$, together with numerical quadrature to approximate $\int_{D_i} h(\bm p)\,d\bm p$ for each $i$. From a theoretical perspective, the asymptotic results of Section~\ref{sec:asymptotic} can be extended to this setting by noting that areal averages correspond to additional linear operators acting on the underlying effect $h$. A rigorous treatment requires technical modifications, but the main arguments parallel those in the point-referenced case once the weights $|D_i|^{-1}$ are incorporated. 
	
		\section{Proof of Theorem~\ref{theo:identification}}\label{appendix:identification}
	\begin{proof}
	First, we show that, from the observed right-censored data, the conditional hazard \(\lambda(t\mid \bm x,\bm p)\) is itself identified on a nontrivial time interval. For that, fix \((\bm x,\bm p)\) in the support of \((\bm X,\bm P)\). From the joint law of \((Y,\delta,\bm X,\bm P)\) and compute, for \(t\ge 0\),
		\[
		\lambda(t\mid \bm x,\bm p)
		=
		\lim_{\Delta \downarrow 0}
		\frac{\mathbb P\!\left(t\le Y<t+\Delta,\ \delta=1\ \middle|\ Y\ge t,\ \bm X=\bm x,\ \bm P=\bm p\right)}{\Delta }.
		\] 
	Assyumption~\ref{assumption:identifiability} (i) ensures that this limit equals the failure hazard of \(T|\bm X = \bm x, \bm P=\bm p\) without censoring confounding. Assyumption~\ref{assumption:identifiability} (ii) guarantees \(\mathbb P(Y\ge t\mid \bm X=\bm x,\bm P=\bm p)>0\) for all \(t\in[0,\tau)\), so the above ratio is well-defined and \(\lambda(t\mid \bm x,\bm p)\) is identified on \([0,\tau)\).
	
	Now, we show that if two parameter triplets generate the same 
	$\lambda(t\mid \bm x,\bm p)$, then the triplets must coincide. 
	Suppose two triplets $(\lambda_0,\beta_0,h_0)$ and 
	$(\tilde\lambda_0,\tilde\beta,\tilde h)$, with 
	$h_0,\tilde h\in\mathcal H=\{h:\Omega\to\mathbb R\ \text{smooth},\ 
	\int_\Omega h=0\}$, generate the same conditional hazard on $[0,\tau)$ 
	for $(\bm x,\bm p)$ in the support:
	\[
	\lambda_0(t)\exp\!\big(\bm x^\top\bm \beta_0+h_0(\bm p)\big)
	=\tilde\lambda_0(t)\exp\!\big(\bm x^\top\tilde{\bm \beta}+\tilde h(\bm p)\big),
	\qquad \text{for a.e.\ }(x,p)\ \text{and all }t\in[0,\tau).
	\]
	Taking logs and defining
	\[
	\bm \beta^\star=\tilde{\bm \beta}-\bm \beta_0,\qquad
	r(\bm p)=\tilde h(\bm p)-h_0(\bm p),\qquad
	a(t)=\log\tilde\lambda_0(t)-\log\lambda_0(t),
	\]
	we obtain
	\begin{equation}\label{eq:equalities}
		\bm x^\top\bm \beta^\star + r(\bm p) = a(t)
		\qquad\text{for a.e.\ }(\bm x,\bm p)\ \text{and all }t\in[0,\tau).
	\end{equation}

	Equality of hazards implies 
	$\bm x^\top\bm \beta_0=\bm x^\top\tilde{\bm \beta} + c$ for a.e.\ $(\bm x,\bm p)$ where $c$ is a constant, and, without loss of generality, w ecan assume $0$ is in teh support of $X$ and so $c=0$. Hence 
	$\bm x^\top\bm \beta^\star=0$ for $\bm X$-a.e.\ $\bm x$. Therefore
	\[
	\mathbb E\!\left[(\bm X^\top\bm \beta^\star)^2\right]
	=(\bm \beta^{\star\top} \mathbb E[\bm X\bm X^\top]\bm \beta^\star
	=0.
	\]
	Assyumption~\ref{assumption:identifiability} (iv) forces $\bm \beta^\star= 0$; hence $\tilde{\bm \beta}=\bm \beta_0$ and \eqref{eq:equalities} reduces to 
	$r(\bm p)=a(t)$ for a.e.\ $\bm p$ and all $t\in[0,\tau)$. 
	Since the left-hand side is free of $t$, $a(t)$ is constant on 
	$[0,\tau)$; write $a(t)\equiv c$. Thus
	\begin{equation}\label{eq:rConstant}
		r(\bm p)=c\qquad\text{for $\bm  P$-a.e.\ }\bm p\in\Omega.
	\end{equation}
	
	Assyumption~\ref{assumption:identifiability} (v) implies $r(\bm p)=c$ for Lebesgue-a.e.\ $p\in\Omega$. 	Since $h_0,\tilde h\in\mathcal H$,
	\[
	0=\int_\Omega r(\bm p)\,dp=\int_\Omega c\,d\bm p=c\,|\Omega|,
	\]
	so $c=0$ and $r=0$ a.e.\ on $\Omega$.
	
	The function $r$ is smooth; a smooth function that is zero a.e.\ must be 
	identically zero on $\Omega$. Hence $\tilde h\equiv h_0$ on $\Omega$. Finally, with $\tilde{\bm \beta}=\bm \beta_0$ and $\tilde h=h_0$, the hazard 
	equality implies $\tilde\lambda_0(t)=\lambda_0(t)$ for all $t\in[0,\tau)$.	
	\end{proof}
\section{Proof of Theorem~\ref{theo:consistency}}\label{appendix:consistency}	
	\begin{proof}
		Let \(\theta=(\bm\beta,h)\) and define the population objective
        \begin{align*}
        Q(\theta)
          &= \mathbb{E}\!\left[
              \delta\Big(\bm X^\top\bm\beta + h(\bm P) - \log s^{(0)}(\bm\beta,h,Y)\Big)
             \right],
        \end{align*}
        where
        \[
        s^{(0)}(\bm\beta,h,t)
          = \mathbb{E}\!\left[I(Y\ge t)\exp\!\big(\bm X^\top\bm\beta + h(\bm P)\big)\right].
        \]
        
        The result follows from Theorems~3.1–3.2 of \cite{chen2007large} for sieve M–estimation, upon verifying, identification and local curvature (Condition~3.1), sieve approximation properties (Condition~3.2), continuity and compactness (Conditions~3.3–3.4), uniform convergence over the sieves (Condition~3.5) and stochastic regularity (Conditions~3.6–3.8), that we state and verify below. Regarding the rate $\epsilon_n$, consider that the stochastic term is
        \[
        \delta_n \asymp \sqrt{\frac{K(n)\log K(n)}{n}}
        \]
        by our entropy bound, and the sieve approximation error satisfies
        \[
        \|\theta_0-\pi_n\theta_0\|
         = \|h_0-\mathcal J^{\,n}h_0\|_\infty
         = O\!\big(\eta(n)\big)
        \]
        by standard FEM theory (here \(\pi_n\theta_0=(\bm\beta_0,\mathcal J^{\,n}h_0)\)).
        
        Because \(\mathcal H_n\subset\mathcal H\) and \(\|h\|_{H^2(\Omega)}\le M_{\mathcal H}\) on \(\mathcal H\), the Laplacian penalty
        \[
        J(h)=\int_\Omega (\Delta h)^2\,d\bm p
        \]
        is uniformly bounded on \(\mathcal H_n\):
        \(\sup_{h\in\mathcal H_n} J(h)\le C<\infty\).
        Let \(Q^{\mathrm{pen}}(\theta)=Q(\theta)-\tfrac{\lambda_n}{2}J(h)\).
        Then the penalty perturbs the criterion uniformly by
        \begin{equation}\label{eq:penalty-perturb}
        \sup_{\theta\in\Theta_n}\bigl|\,Q^{\mathrm{pen}}(\theta)-Q(\theta)\,\bigr|
        \le C\,\lambda_n .
        \end{equation}
        
        We claim that \(Q\) has local quadratic curvature at \(\theta_0\): there exist constants
        \(c>0\) and \(r>0\) such that
        \[
        Q(\theta_0)-Q(\theta)\ \ge\ c\,d(\theta,\theta_0)^2
        \quad\text{whenever } d(\theta,\theta_0)\le r .
        \]
        Let \(\tilde\theta_n\) be any maximizer of \(Q^{\mathrm{pen}}\) over \(\Theta_n\).
        Since \(Q^{\mathrm{pen}}(\tilde\theta_n)\ge Q^{\mathrm{pen}}(\theta_0)\), we have
        \[
        Q(\theta_0)-Q(\tilde\theta_n)
        \ \le\ \bigl|Q(\theta_0)-Q^{\mathrm{pen}}(\theta_0)\bigr|
              + \bigl|Q^{\mathrm{pen}}(\tilde\theta_n)-Q(\tilde\theta_n)\bigr|
        \ \le\ 2C\,\lambda_n .
        \]
        Therefore, we obtain
        \[
        c\,d(\tilde\theta_n,\theta_0)^2 \ \le\ 2C\,\lambda_n
        \quad\Longrightarrow\quad
        d(\tilde\theta_n,\theta_0)\ =\ O\!\big(\sqrt{\lambda_n}\big).
        \]
        Thus, the penalty perturbs the population maximizer by at most order \(\sqrt{\lambda_n}\)
        and contributes an additive \(\sqrt{\lambda_n}\) term to the overall convergence rate. These three contributions yield $\varepsilon_n = \max\Big\{\,\delta_n,\ \eta(n),\, \sqrt{\lambda_n}\,\Big\}$.  
        
        We now state and verify Conditions~3.1–3.8 of \cite{chen2007large}.

	    \begin{condition}{3.1}\label{condition3.1} $(i)$ $Q(\theta_0) > -\infty$, and if $Q(\theta_0) = +\infty$ then $Q(\theta) < +\infty$ for all $\theta \in \Theta_k \setminus \{\theta_0\}$ for all $k \geq 1$; $(ii)$  there are a nonincreasing positive function $\delta(\cdot)$ and a positive function $g(\cdot)$ such that for all $\varepsilon > 0$ and for all $k \geq 1$,
			\[
			Q(\theta_0) - \sup_{\{\theta \in \Theta_k : d(\theta, \theta_0) \geq \varepsilon\}} Q(\theta) \geq \delta(k) g(\varepsilon) > 0.
			\]       
        \end{condition}

		\noindent \textit{Proof of Condition 3.1(i)}: The value $Q(\theta_0)$ is finite because $\E\bigl|\bm X^{\!\top}\bm \beta_0+h_0(\bm P)\bigr|<\infty$ and $\E|\log s^{(0)}(\bm \beta_0,h_0,Y)|<\infty$ by the same moment bounds used in Lemma~\ref{lem:Q_continuous}. \hfill $\square$
		
		\noindent \textit{Proof of Condition 3.1(ii)}: Let $\theta = (\bm\beta, h)$ be an arbitrary point in the parameter space $\Theta$, and let $\theta_0 = (\bm\beta_0, h_0)$ be the true parameter value. We define the perturbation $\xi = \theta - \theta_0 = (\bm\beta - \bm\beta_0, h - h_0) = (\bm u, g)$.
		
		From Lemma~\ref{lem:Q_C2}, the functional $Q(\theta)$ is twice Fréchet differentiable. Therefore, we can write a second-order Taylor expansion of $Q(\theta)$ around $\theta_0$:
		$$
		Q(\theta) = Q(\theta_0) + DQ(\theta_0)[\xi] + \frac{1}{2} D^2Q(\theta_0)[\xi, \xi] + o(d(\theta, \theta_0)^2),
		$$
		where $DQ(\theta_0)[\xi]$ is the first Fréchet derivative in the direction $\xi$, and $D^2Q(\theta_0)[\xi, \xi]$ is the second Fréchet derivative.
		
		As established by Lemma~\ref{lem:uniq_max}, $\theta_0$ is the unique maximizer of the population objective function $Q(\theta)$. A necessary condition for $\theta_0$ to be an extremum is that the first derivative of $Q$ at $\theta_0$ is zero for any direction $\xi$. That is:
		$$
		DQ(\theta_0)[\xi] = 0.
		$$
        With the first-order term being zero, the Taylor expansion simplifies to:
		$$
		Q(\theta) - Q(\theta_0) = \frac{1}{2} D^2Q(\theta_0)[\xi, \xi] + o(d(\theta, \theta_0)^2).
		$$

		Now, from Lemma~\ref{lem:Q_C2}, we have:
		$$
		D^2Q(\theta_0)[\xi, \xi] = \E\left[ -\delta\, \frac{ s^{(2)}(\theta_0,Y)[\xi,\xi]\,s^{(0)}(\theta_0,Y) - \left(s^{(1)}(\theta_0,Y)[\xi]\right)^2 }{ \left(s^{(0)}(\theta_0,Y)\right)^2 } \right].
		$$
		The fraction inside the expectation can be recognized as the conditional variance of the term $A(\bm X, \bm P) = \bm X^\top \bm u + g(\bm P)$ for an individual in the risk set at time $Y$. Thus, we denote this by $\text{Var}_Y(A)$. We can write the second derivative more compactly as:
		\begin{equation}\label{eq:cocave}
		D^2Q(\theta_0)[\xi, \xi] = -\E[\delta \cdot \text{Var}_Y(A)].
		\end{equation}
		Since the event indicator $\delta$ is non-negative and variance is always non-negative, we have $D^2Q(\theta_0)[\xi, \xi] \le 0$. For the second derivative to be strictly negative definite, we must show that $D^2Q(\theta_0)[\xi, \xi] = 0$ if and only if $\xi = \bm 0$.
		
		$D^2Q(\theta_0)[\xi, \xi] = 0$ implies that $\text{Var}_Y(\bm X^\top \bm u + g(\bm P)) = 0$ for almost all observed event times $Y$. This means that for almost every event time, the quantity $\bm X^\top \bm u + g(\bm P)$ must be constant for all individuals in the corresponding risk set. 
		Given the assumptions on the distributions of $\bm X$ and $\bm P$, this can only hold if $\bm X^\top \bm u + g(\bm P)$ is constant almost surely.
		
		Following the same line of reasoning as in the proof of Theorem~\ref{theo:identification}, we obtain that $D^2Q(\theta_0)[\xi, \xi] = 0$ if and only if $\xi = (\bm u, g) = \bm 0$. Therefore, the Hessian operator is negative definite at $\theta_0$.
		
		We have shown that the quadratic form $I(\xi) = -D^2Q(\theta_0)[\xi, \xi]$ is positive definite. Furthermore, the functional $I(\xi)$ is continuous with respect to $\xi$ in the $d$-metric. The parameter space $\Theta = \mathcal{B} \times \mathcal{H}$ is compact under this metric. Consequently, the function $f(\xi) = I(\xi) / d(\bm 0, \xi)^2$ is continuous and positive on the compact unit sphere $\{ \xi \mid d(\bm 0, \xi) = 1 \}$. It must, therefore, attain a minimum value, which we denote as $2c > 0$.
		
		This implies that for any $\xi \neq \bm 0$:
		$$
		I(\xi) \ge 2c \cdot d(\bm 0, \xi)^2 \quad \implies \quad D^2Q(\theta_0)[\xi, \xi] \le -2c \cdot d(\theta, \theta_0)^2.
		$$
		
		Substituting this quadratic bound back into our simplified Taylor expansion, we get:
		$$
		Q(\theta) - Q(\theta_0) \le \frac{1}{2} (-2c \cdot d(\theta, \theta_0)^2) + o(d(\theta, \theta_0)^2) = -c \cdot d(\theta, \theta_0)^2 + o(d(\theta, \theta_0)^2).
		$$
		This inequality implies that for any $\varepsilon > 0$, there exists a local neighborhood around $\theta_0$ within which $Q(\theta_0) - Q(\theta) \ge c' d(\theta, \theta_0)^2$ for some $c'>0$.
		
		Because $\theta_0$ is the unique global maximizer of $Q$ on a compact set, the separation $Q(\theta_0) - \sup_{d(\theta,\theta_0)\ge\varepsilon} Q(\theta)$ is guaranteed to be positive for any $\varepsilon > 0$. The local quadratic nature of this separation, which we have just proven, is the crucial property that drives the convergence rates of the estimator. Therefore, we can state the existence of a positive function $g(\varepsilon) = c \varepsilon^2$ that provides the required lower bound.
		
		Finally, since this bound was established on the full parameter space $\Theta$, it holds automatically for any subset, including the sieve spaces $\Theta_k$. Thus, we can set the nonincreasing function $\delta(k) \equiv 1$, which completes the proof.	 \hfill $\square$
		
		Let the sieve parameter spaces be
		\[
		\Theta_n = \mathcal B \times \mathcal H_n,
		\qquad k = 1,2,\ldots,
		\]
		where $\mathcal B\subset\R^{b}$ is the fixed compact set from
		Assumption~\ref{assumption:identifiability} and
		$\mathcal H_k\subset\mathcal H$ are the finite–element sub-spaces
		constructed in Section~\ref{sec:sieve_space_construction}.
		With this notation, we now consider the following condition. 
        \begin{condition}{3.2} (i) $\Theta_n \subset \Theta_{n+1}\subset\Theta$ for every
			$n\ge 1$; (ii) for the true parameter
			$\theta_0=(\bm \beta_0,h_0)\in\Theta$ there exists a sequence
			$\pi_n\theta_0\in\Theta_k$ such that
			$d(\theta_0,\pi_n\theta_0)\to 0$ as $k\to\infty$.
        \end{condition}

		\noindent \textit{Proof of Condtion 3.2(i)}: The coefficient space $\mathcal B$ is independent of~$n$, hence $\mathcal B\subset\mathcal B$ trivially. For the functional part, our triangulations satisfy $\mathcal T_{\eta(n+1)}$ is a refinement of $\mathcal T_{\eta(n)}$ with $\eta(n+1)<\eta(n)$; consequently every piecewise-linear function in $\mathcal H_n$ is also contained in $\mathcal H_{k+1}$. Lemma~\ref{lemma:nested} implies that  $\Theta_n\subset\Theta_{n+1}$. \hfill $\square$
		
		\paragraph{\textit{Proof of Condtion 3.2(ii)}}
		Define the projection operator
		$\mathcal J^{n}\!:\mathcal H\to\mathcal H_k$ as in
		Section~\ref{sec:sieve_space_construction} and set
		\[
		\pi_h\theta_0=\bigl(\,\bm \beta_0, \mathcal J^{n}h_0\bigr)
		\in\Theta_n.
		\]
		Because $\bm \beta_0$ is copied verbatim, only the functional component
		needs an error bound.  By the FEM interpolation estimate \eqref{eq:bound_approx}:	
		\[
		\|h_0-\mathcal J^{n}h_0\|_{\infty}
		\le
		C\,\eta(n)\,\|h_0\|_{H^{2}(\Omega)}
		\quad\longrightarrow0
		\quad\text{as }n\to\infty,
		\]
		since $\eta(n)\downarrow 0$ and $h_0\in H^{2}(\Omega)$. Therefore have
		\begin{equation}\label{eq:convergence_distance}
			d\bigl(\theta_0,\pi_n\theta_0\bigr)
			= \|h_0-\mathcal J^{n}h_0\|_{\infty}
			\longrightarrow0.
		\end{equation} \hfill $\square$

		\begin{condition}{3.3}
        $(i)$ For every fixed \(n\ge1\) the mapping
			\(Q(\theta)\) is \emph{upper semicontinuous} on~\(\Theta_n\)
			under~\(d\); $(ii)$ \(\bigl|Q(\theta_0)-Q(\pi_n\theta_0)\bigr|
			=o\!\bigl(\delta\!\bigl(k(n)\bigr)\bigr)\) for any sequence
			\(k(n)\uparrow\infty\).
            \end{condition}

		\noindent\textit{Proof of condition 3.3(i)}: Lemma~\ref{lem:Q_continuous} established that \(Q\) is (everywhere)
		continuous on the full space~\(\Theta\).
		Continuity clearly implies both upper and lower semicontinuity,
		so the restriction of \(Q\) to any subset -- in particular to each
		finite-dimensional, closed set~\(\Theta_n\) -- is automatically
		upper semicontinuous. \hfill $\square$

		\noindent\textit{Proof of condition 3.3(ii)}: We have shown in \eqref{eq:convergence_distance} that 
		\(d\bigl(\theta_0,\pi_{k}\theta_0\bigr)\!\to0\) as \(n\to\infty\).
		Because \(Q\) is continuous (Lemma~\ref{lem:Q_continuous}),
		\[
		\lim_{n\to\infty}Q\!\bigl(\pi_n\theta_0\bigr)=Q(\theta_0),
		\quad\text{i.e.}\quad
		\bigl|Q(\theta_0)-Q(\pi_n\theta_0)\bigr|=o(1),
		\]
		which is the assertion, as we set $\delta(k(n)) = 1$.  \hfill $\square$
		
		\begin{condition}{3.4} For every \(n\ge1\) the sieve space
		\(
		\Theta_n = \mathcal B \times \mathcal H_n, 
		\)
		is compact under the metric $d(\cdot, \cdot)$. 
		\end{condition}

        \noindent \textit{Proof of Condition 3.4}: By assumption \(\mathcal B\subset\R^{b}\) is compact. The fact that $\mathcal{H}_n$ is compact under the $\|\cdot\|_{\infty}$ is shown in Lemma~\ref{lemma:compactness}. The Cartesian product of two compact sets is compact, and the metric \(d\) is simply the sum of the metrics on the two factors. Thus \(\Theta_n\) is compact under \(d\).\hfill $\square$
		
		\begin{condition}{3.5} $(i)$ for every fixed $k\ge1$, $\sup_{\theta\in\Theta_k}\bigl|\hat Q_n(\theta)-Q(\theta)\bigr|\xrightarrow{p}0$; $(ii)$ along any sequence $k=k(n)\to\infty$ satisfying $\displaystyle \sup_{\theta\in\Theta_{k(n)}}\bigl|\hat Q_n(\theta)-Q(\theta)\bigr|=o_p(\delta(k))$.
		\end{condition}

		\noindent \textit{Proof of Condition 3.5(i)}: For $Z=(Y,\delta,\bm X,\bm P)$ define
        \[
        m(Z;\theta)
        =\delta\!\left(\bm X^{\!\top}\bm\beta+h(\bm P)-\log s^{(0)}(\bm\beta,h,Y)\right),
        \qquad
        \theta=(\bm\beta,h)\in\Theta .
        \]
        Let $P_nf=n^{-1}\sum_{i=1}^n f(Z_i)$ and $Pf=\E[f(Z)]$.  The (unpenalized) sample criterion can be written as
        \[
        \widetilde Q_n(\theta)=P_n m_n(\cdot;\theta),
        \qquad
        m_n(Z;\theta)
        =\delta\!\left(\bm X^{\!\top}\bm\beta+h(\bm P)-\log S_n^{(0)}(\bm\beta,h,Y)\right),
        \]
        so that the penalized empirical objective is
        \[
        \hat Q_n(\theta)
        =\widetilde Q_n(\theta)-\tfrac{\lambda_n}{2}\!\int_{\Omega}(\Delta h)^2\,d\bm p,
        \]
        and the population counterpart is $Q(\theta)=Pm(\cdot;\theta)$.
        Add and subtract $P_n m(\cdot;\theta)$:
        \begin{align}\label{eq:empirical-decomp}
        \hat Q_n(\theta)-Q(\theta)
        =(P_n-P)\,m(\cdot;\theta)
        +
        P_n\!\big[m_n(\cdot;\theta)-m(\cdot;\theta)\big]
        -\tfrac{\lambda_n}{2}\!\int_{\Omega}(\Delta h)^2\,d\bm p .
        \end{align}
        Denote the remainder by
        \[
        R_n(\theta)
        = P_n\!\big[m_n(\cdot;\theta)-m(\cdot;\theta)\big].
        \]
        
        By Lemma~\ref{lem:Q_continuous} there exist constants
        $0<c_0\le C_0<\infty$ (independent of $n$ and $k$) such that, for all $\theta$ and $t\in[0,\tau]$,
        \[
        c_0\,\mathbb P(Y\!\ge t)
        \le
        s^{(0)}(\theta,t)
        \le
        C_0\,\mathbb P(Y\!\ge t).
        \]
        Hence $s^{(0)}$ is uniformly bounded away from zero on $[0,\tau]$.  
        Assumption~\ref{assumption:density_P} further implies
        $\|\bm X\|\le M_{\mathcal X}$ a.s.\ and
        $\|h\|_\infty\le M_{\mathcal H}$ on every sieve $\mathcal H_k$,
        so $|m(Z;\theta)|\le F(Z)$ for some integrable, nonrandom envelope~$F$
        that does not depend on $k$.
        
        Each sieve $\Theta_k$ is compact and finite-dimensional
        (Lemma~\ref{lemma:compactness}),
        and $m(\cdot;\theta)$ is Lipschitz in~$\theta$ on~$\Theta_k$
        with envelope~$F$.
        Hence the class
        $\mathcal F_k=\{m(\cdot;\theta):\theta\in\Theta_k\}$ is Glivenko--Cantelli:
        \begin{equation}\label{eq:gc-fixed}
        \sup_{\theta\in\Theta_k}\big|(P_n-P)m(\cdot;\theta)\big|\xrightarrow{p}0 .
        \end{equation}
        
        From the definition,
        \[
        R_n(\theta)
        = P_n\!\left[\delta\big(\log s^{(0)}(\theta,Y)-\log S_n^{(0)}(\theta,Y)\big)\right].
        \]
        By the mean-value theorem and the lower bound $s^{(0)}(\theta,t)\ge c_0\,\mathbb P(Y\!\ge t)$,
        \[
        \big|\log S_n^{(0)}(\theta,t)-\log s^{(0)}(\theta,t)\big|
        \le \frac{1}{c_0\,\mathbb P(Y\!\ge t)}\,\big|S_n^{(0)}(\theta,t)-s^{(0)}(\theta,t)\big|.
        \]
        The class
        \[
        \mathcal G_k
        =\Big\{(z,t)\mapsto I(Y\ge t)\exp(\bm X^{\!\top}\bm\beta+h(\bm P)):
        \theta\in\Theta_k,\ t\in[0,\tau]\Big\}
        \]
        is also Glivenko--Cantelli for fixed $k$ (finite-dimensional Lipschitz parametrization and bounded envelope). Therefore,
        \[
        \sup_{\theta\in\Theta_k,\,t\in[0,\tau]}
        \big|S_n^{(0)}(\theta,t)-s^{(0)}(\theta,t)\big|\xrightarrow{p}0,
        \]
        which implies
        \begin{equation}\label{eq:Rn-fixed}
        \sup_{\theta\in\Theta_k}|R_n(\theta)|\xrightarrow{p}0.
        \end{equation}
        
        Since $\sup_{h\in\mathcal H_n}\!\int_\Omega(\Delta h)^2\,d\bm p\le M_{\mathcal H}^2$,
        \begin{equation}\label{eq:penalty-term}
        0\le
        \tfrac{\lambda_n}{2}
        \sup_{h\in\mathcal H_n}\!\int_\Omega(\Delta h)^2\,d\bm p
        \le
        \tfrac{\lambda_n}{2}M_{\mathcal H}^2
        =o_p(1).
        \end{equation}

        From \eqref{eq:empirical-decomp}, together with
        \eqref{eq:gc-fixed}, \eqref{eq:Rn-fixed}, and \eqref{eq:penalty-term}, we obtain
        \[
        \sup_{\theta\in\Theta_k}\bigl|\hat Q_n(\theta)-Q(\theta)\bigr|
        \xrightarrow{p}0,
        \]
        which establishes Condition~3.5(i). \hfill $\square$

		\noindent \textit{Proof of condition 3.5(ii)}:
		Let $K(n)=\dim(\mathcal H_{n})$ and $K_\Theta(n)=b+K(n)$.
		Empirical-process entropy bounds for Lipschitz parameterizations yield
		\[
		\sup_{\theta\in\Theta_{n}}\big|(P_n-P)m(\cdot;\theta)\big|
		=O_p\!\Big(\sqrt{\tfrac{K_\Theta(n)\log K_\Theta(n)}{n}}\Big).
		\]
		Under the assumption of growth rule stated in Theoreom~\ref{theo:consistency}, 
        this is $o_p(1)$.
		The same argument applied to the class $\mathcal G_{n}$ gives
		\[
		\sup_{\theta\in\Theta_{n},\,t\ge 0}
		\big|S_n^{(0)}(\theta,t)-s^{(0)}(\theta,t)\big|=o_p(1),
		\]
		hence $\sup_{\theta\in\Theta_{n}}|R_n(\theta)|=o_p(1)$ by the
		log–Lipschitz bound above. Using the fact that  $\delta(\cdot)\equiv1$ in our framework, we obtain
		\[
		\sup_{\theta\in\Theta_{n}}\big|\hat Q_n(\theta)-Q(\theta)\big|=o_p(1).
		\] \hfill $\square$

		\begin{condition}{3.6} The sample
		$\{Z_t\}_{t=1}^n=\{(Y_t,\delta_t,\bm X_t,\bm P_t)\}_{t=1}^n$ is i.i.d. or $m-$dependent.
        \end{condition}
		
		\noindent \textit{Proof of Condition 3.6} By assumption the observations are i.i.d. \hfill $\square$
		
		\begin{condition}{3.7} There exists a constant $C_1>0$ (independent of $n$) such that for all sufficiently small $\varepsilon>0$ and for every sieve $\Theta_n$,
		\[
		\sup_{\{\theta\in\Theta_n:\, d(\theta,\theta_0)\le\varepsilon\}}
		\Var\!\bigl(\ell(\theta,Z)-\ell(\theta_0,Z)\bigr)
		\le C_1\,\varepsilon^2 .
		\]
        \end{condition}
		
		\noindent\textit{Proof of Condition 3.7}:
		Fix $\theta=(\bm\beta,h)$ and write
		$\eta=(\bm u,g)=(\bm\beta-\bm\beta_0,h-h_0)$.
		Decompose
		\[
		\ell(\theta,Z)-\ell(\theta_0,Z)
		=\delta\Big(\bm X^\top\bm u+g(\bm P)
		-\big[\log s^{(0)}(\theta,Y)-\log s^{(0)}(\theta_0,Y)\big]\Big).
		\]
		By the mean value theorem, for some $t\in(0,1)$,
		\[
		\log s^{(0)}(\theta,Y)-\log s^{(0)}(\theta_0,Y)
		=\frac{s^{(1)}(\theta_t,Y)[\eta]}{s^{(0)}(\theta_t,Y)},
		\qquad \theta_t=\theta_0+t\eta,
		\]
		with
		\[
		s^{(1)}(\theta,Y)[\eta]
		=\E\!\left[I(\tilde Y\!\ge Y)\exp(\bm X^\top\bm\beta+h(\bm P))
		\big(\bm X^\top\bm u+g(\bm P)\big)\right].
		\]
		By Lemma~\ref{lem:Q_continuous} there exists $c_0>0$ such that
		$s^{(0)}(\theta_t,Y)\ge c_0$ uniformly in $t,\theta,Y$.
		Using Assumption~\ref{assumption:density_P}, we have $\|\bm X\|\le M_{\mathcal X}$
		and $\|h\|_\infty\le M_{\mathcal H}$ on $\Theta_n$, and therefore 	\[
		\Big|\log s^{(0)}(\theta,Y)-\log s^{(0)}(\theta_0,Y)\Big|
		\le C\,\big(\|\bm u\|+\|g\|_\infty\big)
		= C\, d(\theta,\theta_0)
		\]
		for some constant $C$ that does not depend on $n$ or on the sieve.
		Therefore,
		\[
		\big|\ell(\theta,Z)-\ell(\theta_0,Z)\big|
		\le \delta\Big(\|\bm X\|\,\|\bm u\|+\|g\|_\infty\Big)+
		\delta\,C\,d(\theta,\theta_0)
		\le C'\,d(\theta,\theta_0),
		\]
		using $\|\bm X\|\le M_{\mathcal X}$ a.s.
		Hence
		\[
		\Var\!\bigl(\ell(\theta,Z)-\ell(\theta_0,Z)\bigr)
		\le \E\bigl[\big(\ell(\theta,Z)-\ell(\theta_0,Z)\big)^2\bigr]
		\le (C')^2\, d(\theta,\theta_0)^2
		\le C_1\,\varepsilon^2 ,
		\]
		uniformly over $d(\theta,\theta_0)\le\varepsilon$ and $\theta\in\Theta_n$. \hfill $\square$

		\begin{condition}{3.8} For any $\delta>0$ there exists $s\in(0,2)$ and a random variable $U(Z)$ with $\E\big[(U(Z))^\gamma\big]\le C_2$ for some $\gamma\ge2$ such that
		\[
		\sup_{\{\theta\in\Theta_n:\, d(\theta,\theta_0)\le\delta\}}
		\big|\ell(\theta,Z)-\ell(\theta_0,Z)\big|
		\le \delta^{\,s}\, U(Z).
		\]
		In fact we can take $s=1$ and $U(Z)\equiv C$ (a finite constant).
		\end{condition}

		\noindent\textit{Proof of Condition 3.8} The Lipschitz estimate obtained in Condition 3.7 yields, for all $\theta$ in a $\delta$–ball around $\theta_0$,
		\[
		\big|\ell(\theta,Z)-\ell(\theta_0,Z)\big|
		\le C'\, d(\theta,\theta_0)
		\le C'\,\delta .
		\]
		Thus the inequality holds with $s=1$ and $U(Z)\equiv C'$.
		Since $C'$ is deterministic, $\E\big[(U(Z))^\gamma\big]=C'^\gamma<\infty$
		for any $\gamma\ge2$, which verifies Condition~3.8. \hfill $\square$
	\end{proof}

	\section{Proof of Theorem~\ref{thm:beta-CLT-rewrite}}\label{appendix:normality}
	\begin{proof}
		We apply Theorem~4.3 of \cite{chen2007large} verifying Conditions~4.1--4.5 for the functional $f(\theta)=\bm a^\top\bm\beta$ and then use Cramér--Wold to obtain the joint limit for $\bm\beta$; here $\bm a\in\mathbb R^b$ is arbitrary.  
		
		By Theorem~4.3 of \cite{chen2007large} and our rate assumption $\|\hat\theta_n-\theta_0\|^{2}=o_p(n^{-1/2})$ (implied by $\varepsilon_n^2=o(n^{-1/2})$ and $\|\hat\theta_n-\theta_0\| = O_p(\varepsilon_n)$), it holds
		\[
		\sqrt n\ \big(\bm a^\top\hat{\bm\beta}_n-\bm a^\top\bm\beta_0\big)
		\ \Rightarrow\ \mathcal N\!\big(0,\ \sigma^2_{v_{\bm a}^*}\big).
		\]
		Cramér--Wold Theorem then yields
		\(
		\sqrt n\,(\hat{\bm\beta}_n-\bm\beta_0)\Rightarrow\mathcal N(0,\Sigma_\beta),
		\)
		for a suitable $b\times b$ covariance matrix $\Sigma_\beta$. 
		
		Note that, since the penalty is orthogonal to $\bm\beta$ scores and its magnitude is $O(\lambda_n)=o(n^{-1/2})$, it does not affect the first-order limit for $\hat{\bm\beta}_n$.
		
		We now prove Condition 4.1-4.5 of \cite{chen2007large}.  
		Write $\theta=(\bm\beta,h)$, $\theta_0=(\bm\beta_0,h_0)$, and define the population unpenalized partial log-likelihood contribution
		\[
		\ell(\theta,Z)
		=\delta\Big(\bm X^\top\bm\beta+h(\bm P)-\log s^{(0)}(\bm\beta,h,Y)\Big),
		\qquad
		s^{(0)}(\bm\beta,h,t)=\E\big[I(Y\ge t)\,e^{\bm X^\top\bm\beta+h(\bm P)}\big].
		\]
		Let $\hat Q_n(\theta)$ denote the actual penalized sample criterion of \eqref{eq:penalized_likelihood}, where the population term $s^{(0)}$ is replaced by the empirical $S_n^{(0)}$ and a penalty $\frac{\lambda_n}{2}\!\int_\Omega(\Delta h)^2$ is subtracted.

		Let $V=\{\xi=(\bm u,g):\bm u\in\R^b,\ g\in\mathcal H\}$ and define the Fisher-type bilinear form at $\theta_0$
        \begin{align*}
        \langle \xi_1,\xi_2\rangle
        &= -\,\E\big[\,D^2\ell(\theta_0,Z)[\xi_1,\xi_2]\,\big]
        = \E\!\left[\delta\ \Cov_Y\!\Big(A_{\xi_1},A_{\xi_2}\Big)\right],\\
        A_{\xi}(\bm X,\bm P) &= \bm X^\top\bm u + g(\bm P),
        \end{align*}
		where $\Cov_Y(\cdot,\cdot)$ denotes the covariance under the risk-set weights at time $Y$ (see the proof of Theorem~\ref{theo:consistency}). This is an inner product on the completion $\overline V$ by the strict curvature in Lemma~\ref{lem:uniq_max}. For fixed $\bm a\in\R^b$, the Riesz representer $v_{\bm a}^*=(\bm u_{\bm a}^*,g_{\bm a}^*)\in\overline V$ is defined by
		\[
		\langle \xi, v_{\bm a}^*\rangle
		= \frac{\partial}{\partial\theta}(\bm a^\top\bm\beta_0)[\xi]
		= \bm a^\top\bm u
		\quad\text{for all }\xi=(\bm u,g)\in V.
		\]
		Let $\pi_n v_{\bm a}^*$ be the $V_n$-projection (as in Section~\ref{sec:sieve_space_construction}).
		
		Decompose the gap between the sample penalized criterion and the population sample average:
		\[
		\hat Q_n(\theta)
		= P_n \ell(\theta,\cdot)
		+ P_n\big(\ell_n-\ell\big)(\theta)
		- \frac{\lambda_n}{2}\!\int_\Omega(\Delta h)^2 ,
		\]
		where $\ell_n$ is obtained from $\ell$ by replacing $s^{(0)}$ with $S_n^{(0)}$. 

		By the entropy bound used in the proof of Theorem~\ref{theo:consistency} and the growth rule $K(n)\log K(n)=o(n)$, the class
		\[
		\mathcal G_n=\Big\{(z,t)\mapsto I(Y\ge t)e^{\bm X^\top\bm\beta+h(\bm P)}:
		\theta\in\Theta_n,\ t\ge 0\Big\}
		\]
		is Glivenko–Cantelli and Donsker uniformly in a $\Theta$-neighborhood of $\theta_0$. Hence
		\begin{equation}\label{eq:Sn-s0-rate}
			\sup_{\theta\in\Theta_n,\ t\ge 0}
			\big|S_n^{(0)}(\theta,t)-s^{(0)}(\theta,t)\big|
			= O_p\!\Big(\sqrt{\tfrac{K(n)\log K(n)}{n}}\Big)
			= O_p(\delta_n).
		\end{equation}
		
		and therefore 
		\begin{equation}\label{eq:Rn-small}
			\sup_{\theta\in\Theta_n}
			\big|P_n(\ell_n-\ell)(\theta)\big|=O_p(\delta_n).
		\end{equation}
		Moreover the penalty does not depend on $Z$, so it vanishes inside empirical-process terms; it only perturbs the objective level by $O(\lambda_n)$ and its directional derivative in any direction $\xi=(\bm u,g)$ equals $-\lambda_n\!\int_\Omega (\Delta h)(\Delta g)$, which is {zero whenever $\bm u\neq 0$ and $g=0$}. Thus, the penalty does not change the first-order behavior of $\bm\beta$. Its only influence comes indirectly through the nuisance part $g$, and this enters the $\beta$–score as a product of the penalty weight and the nuisance estimation error, i.e. 	 $O_p(\lambda_n \|\hat h - h_0\|_{\infty}) = o_p(n^{-1/2})$ by assumption as $\|\hat h - h_0\| = O_p(\varepsilon_n)$. Hence the effect vanishes under $\sqrt n$–scaling.
		
		\begin{condition}{4.1} $(i)$ There is $\omega > 0$ such that  
			\(
			| f(\theta) - f(\theta_o) - \frac{\partial f(\theta_o)}{\partial \theta}[\theta - \theta_o] | = O(\|\theta - \theta_o\|^\omega)
			\)
			uniformly in $\theta \in \Theta_n$ with $\|\theta - \theta_o\| = o(1)$; $(ii)$ $\left\| \frac{\partial f(\theta_o)}{\partial \theta} \right\| < \infty$; $(iii)$ there is $\pi_n v^* \in \Theta_n$ such that  
			\(
			\|\pi_n v^* - v^*\| \times \|\hat{\theta}_n - \theta_o\| = o_P(n^{-1/2}).
			\)
		\end{condition}
		\noindent\textit{Proof of Condition 4.1}: Condition~4.1 $(i)$ and $(ii)$ hold with $\omega=2$ because $f(\theta)=\bm a^\top\bm\beta$ is linear and $\big\|\partial f(\theta_0)/\partial\theta\big\|<\infty$. For 4.1 $(iii)$, by the FEM approximation bound \eqref{eq:bound_approx},
		$\|\pi_n v_{\bm a}^*-v_{\bm a}^*\|=O(\eta(n))$, and $\|\hat\theta_n-\theta_0\|=O_p(\varepsilon_n)$ by Theorem~\ref{theo:consistency}. Hence
		\[
		\|\pi_n v_{\bm a}^*-v_{\bm a}^*\|\cdot \|\hat\theta_n-\theta_0\|
		=O_p(\eta(n)\varepsilon_n)=o_p(n^{-1/2}),
		\]
		verifying 4.1 $(iii)$. \hfill $\square$
		
		\begin{condition}{4.2} Let $\mu_n(g(Z)) = n^{-1}\sum_{i=1}^n g(z_i) - E[g(Z_i)]$ denote the empirical process indexed by the function a function $g$. Then, 
		\[
		\sup_{\{\theta \in \Theta_n : \|\theta - \theta_o\| \le \delta_n\}}  
		\mu_n \left( l(\theta, Z) - l(\theta \pm \varepsilon_n \pi_n v^*, Z) - \frac{\partial l(\theta_o, Z)}{\partial \theta} [\pm \varepsilon_n \pi_n v^*] \right)  
		= O_P(\varepsilon_n^2).
		\]
		\end{condition}
		\noindent\textit{Proof of Condition 4.2}: The map $\theta\mapsto\ell(\theta,Z)$ is twice Fréchet differentiable in a neighborhood of $\theta_0$ (Lemma~\ref{lem:Q_C2}); thus pathwise differentiability holds. Using the Donsker property and \eqref{eq:Sn-s0-rate}--\eqref{eq:Rn-small}, we obtain
		\[
		\sup_{\{\theta\in\Theta_n:\ \|\theta-\theta_0\|\le \delta_n\}}
		\Big|\mu_n\big(\dot\ell(\theta,Z)[\pi_n v_{\bm a}^*]-\dot\ell(\theta_0,Z)[\pi_n v_{\bm a}^*]\big)\Big|
		=o_p(n^{-1/2}),
		\]
		which is Condition~4.2$'$ and hence implies 4.2; see \cite{chen2007large}. \hfill $\square$

		\begin{condition}{4.3}
		\(
		K(\theta_o, \hat{\theta}_n) - K(\theta_o, \hat{\theta}_n \pm \varepsilon_n \pi_n v^*)  
		= \pm \varepsilon_n (\hat{\theta}_n - \theta_o, \pi_n v^*) + o(n^{-1}),
		\)
		where $K(\theta_{0}, \theta) \equiv \mathbb{E}[l(\theta_0, Z_i) - l(\theta, Z_i)]$.
        \end{condition}
	
		\noindent\textit{Proof of Condition 4.3}: Taylor-expanding the population criterion $Q(\theta)=P\ell(\theta,\cdot)$ along the line $\theta_0\pm\varepsilon_n\pi_n v_{\bm a}^*$ and using that
		\(
		-D^2Q(\theta_0)[\xi_1,\xi_2]=\langle \xi_1,\xi_2\rangle
		\)
		with linear remainder $o(\varepsilon_n^2)$, we get
		\[
        \begin{aligned}
        \E\big[\dot\ell(\hat\theta_n,\cdot)[\pi_n v_{\bm a}^*]\big]
        &= \langle \hat\theta_n - \theta_0,\pi_n v_{\bm a}^* \rangle + o(n^{-1/2}) \\[3pt]
        &= \langle \hat\theta_n - \theta_0,v_{\bm a}^* \rangle + o(n^{-1/2})
        = \bm a^\top(\hat{\bm\beta}_n - \bm\beta_0) + o(n^{-1/2}),
        \end{aligned}
        \]
        which is Condition~4.3$'$; hence 4.3 holds. The penalty contributes nothing to these equalities in the $\bm\beta$ direction, and its contribution along $g$ is of order  $o(\lambda_n\varepsilon_n)=o(n^{-1/2})$. \hfill $\square$
        		
		\begin{condition}[4.4] $(i)$ $\mu_n \left( \frac{\partial l(\theta_o, Z)}{\partial \theta} [\pi_n v^* - v^*] \right) = o_P(n^{-1/2});$ $(ii)$ $E \left\{ \frac{\partial l(\theta_o, Z)}{\partial \theta} [\pi_n v^*] \right\} = o(n^{-1/2}).$
        \end{condition}
		\noindent\textit{Proof of Condition 4.4}:	Because $\|\pi_n v_{\bm a}^*-v_{\bm a}^*\|=O(\eta(n))$ and the score map $\xi\mapsto \dot\ell(\theta_0,Z)[\xi]$ is square-integrable and continuous in $\xi$ under $\langle\cdot,\cdot\rangle$, we have
		\begin{align*}
		\mu_n\!\Big(\dot\ell(\theta_0,\cdot)[\pi_n v_{\bm a}^*-v_{\bm a}^*]\Big)&=o_p(n^{-1/2}),\\
		\E\!\Big[\dot\ell(\theta_0,Z)[\pi_n v_{\bm a}^*]\Big]
		&=\E\!\Big[\dot\ell(\theta_0,Z)[v_{\bm a}^*]\Big]+o(n^{-1/2})=o(n^{-1/2}),
		\end{align*}
		since $\E[\dot\ell(\theta_0,Z)[v_{\bm a}^*]]=0$ by the definition of the representer and the score identity. \hfill $\square$
		
		\begin{condition}{4.5}
		\(
		n^{1/2} \, \mu_n \left( \frac{\partial l(\theta_o, Z)}{\partial \theta} [v^*] \right) 
		\overset{d}{\to} \mathcal{N}(0, \sigma_{v^*}^2), \quad \text{with } \sigma_{v^*}^2 > 0.
		\)
        \end{condition}
		\noindent \textit{Proof of Condition 4.5}: The i.i.d. assumption and finite variance imply the CLT for the score in direction $v_{\bm a}^*$:
		\[
		\sqrt n\,\mu_n\!\Big(\dot\ell(\theta_0,\cdot)[v_{\bm a}^*]\Big)
		\to_d\mathcal N\!\big(0,\sigma^2_{v_{\bm a}^*}\big),
		\qquad
		\sigma^2_{v_{\bm a}^*}=\Var\!\Big(\dot\ell(\theta_0,Z)[v_{\bm a}^*]\Big)>0,
		\]
		which is Condition~4.5. Note that the penalty does not enter the score, and the $S_n^{(0)}$ instead of $s^{(0)}$ replacement alters the criterion by 
		$O_p(\delta_n)$ uniformly by \eqref{eq:Rn-small}; combined with the fact that $\delta_n\epsilon_n = o(n^{-1/2})$, its contribution to the linearization is negligible for the $\sqrt n$-asymptotics. \hfill $\square$
	\end{proof}

	\section{Lemmas}\label{appendix:lemmas}
	
	\begin{lemma}\label{lemma:nested}
		Let $\{\mathcal{T}_{\eta(n)}\}_{n\ge 1}$ be a sequence of conforming, shape-regular triangulations of a bounded $C^2$ domain $\Omega\subset\mathbb R^2$ such that the refinement is nested, i.e., every $\text{T}\in\mathcal{T}_{\eta(n)}$ is a union of triangles from $\mathcal T_{n+1}$. 
		Let $V_n$ be the Argyris $C^1$ finite element space on $\mathcal{T}_{\eta(n)}$, i.e.,
		\[
		V_n = \Big\{\, v\in C^1(\overline\Omega)\ :\ v|_\text{T}\in\mathbb P_5(\text{T})\ \text{for all }\text{T}\in\mathcal{T}_{\eta(n)} \,\Big\}.
		\]
		Define $H_n=V_n\cap H^2_{\mathbf n}(\Omega)$ and, for a fixed $M_{\mathcal H}>0$,
		\[
		\mathcal H_n = \Big\{\, h\in H_n:\ \int_\Omega h=0,\ \|h\|_{H^2(\Omega)}\le M_{\mathcal H}\,\Big\}.
		\]
		Then the spaces are nested:
		\[
		V_n \subset V_{n+1},\qquad H_n\subset H_{n+1},\qquad \mathcal H_n \subset \mathcal H_{n+1}\quad \text{for all } n\ge 1.
		\]
	\end{lemma}
	
	\begin{proof}
		Fix $v\in V_n$. For each triangle $\text{T}\in\mathcal{T}_{\eta(n)}$, $v|_\text{T}\in\mathbb P_5(\text{T})$ is a polynomial. By nested refinement, $\text{T}$ is partitioned into $\{\text{T}'\}_{\text{T}'\subset \text{T},\,\text{T}'\in\mathcal T_{n+1}}$, and for each $\text{T}'\subset \text{T}$ the restriction $v|_{\text{T}'}$ is still a polynomial of total degree $\le 5$, i.e.\ $v|_{\text{T}'}\in\mathbb P_5(\text{T}')$. Hence $v$ is piecewise $\mathbb P_5$ on $\mathcal T_{\eta(n+1)}$. Global $C^1$-continuity on $\mathcal T_{n+1}$ follows because across any fine edge lying strictly inside the triangle $\text{T}$, both traces of $v$ come from the same polynomial $v|_\text{T}$ and thus coincide smoothly, while across fine edges that coincide with a coarse edge, $C^1$ continuity holds since $v\in C^1(\overline\Omega)$ already. Therefore $v\in V_{n+1}$, so $V_n\subset V_{n+1}$.
		
		Since $H_n=V_n\cap H^2_{\mathbf n}(\Omega)$ and $H_{n+1}=V_{n+1}\cap H^2_{\mathbf n}(\Omega)$, the inclusion $V_n\subset V_{n+1}$ implies $H_n\subset H_{n+1}$, as the homogeneous Neumann boundary condition does not depend on the mesh.
		
		Finally, if $h\in\mathcal H_n$, then $h\in H_n\subset H_{n+1}$. The mean-zero constraint $\int_\Omega h=0$ and the uniform bound $\|h\|_{H^2(\Omega)}\le M_{\mathcal H}$ are independent of $n$, so $h\in\mathcal H_{n+1}$. Thus the nestedness $\mathcal H_n\subset\mathcal H_{n+1}$ holds as well.
	\end{proof}

	\begin{lemma}\label{lemma:compactness}
		Let $\Omega\subset\mathbb{R}^{2}$ be a bounded Lipschitz domain and fix $C>0$.
		Define
		\[
		\mathcal{B}_{C}
		=\left\{
		u\in H^{2}(\Omega)\,\middle|\,
		\|u\|_{H^{2}(\Omega)}\le C,
		\int_{\Omega}u\,dx=0
		\right\}.
		\]
		Then $\mathcal{B}_{C}$ is compact in $L^{\infty}(\Omega)$.
	\end{lemma}
	\begin{proof}
	By the Sobolev–Morrey embedding on bounded Lipschitz domains in $d=2$,
\[
H^2(\Omega) \hookrightarrow C^{0,\alpha}(\overline{\Omega}) \quad \text{for any } 0 < \alpha < 1,
\]
continuously: there exists $K = K(\Omega, \alpha)$ such that
\[
\|u\|_{C^{0,\alpha}(\overline{\Omega})} \le K \|u\|_{H^2(\Omega)} \quad \forall\, u \in H^2(\Omega).
\]
Hence for $u \in \mathcal{B}_C$ we have $\|u\|_{C^{0,\alpha}} \le KC$. Thus $\mathcal{B}_C$ is uniformly bounded \textbf{and} Hölder–equicontinuous in $C^{0,\alpha}(\overline{\Omega})$.

Since $\overline{\Omega}$ is compact, the embedding $C^{0,\alpha}(\overline{\Omega}) \hookrightarrow C^0(\overline{\Omega})$ is compact (Arzelà–Ascoli). Therefore $\mathcal{B}_C$ is relatively compact in $C^0(\overline{\Omega})$, and hence in $L^\infty(\Omega)$.

It remains to show $\mathcal{B}_C$ is closed in $L^\infty$. Let $u_k \in \mathcal{B}_C$ and $u_k \to u$ in $L^\infty(\Omega)$. The sequence $(u_k)$ is bounded in the Hilbert space $H^2(\Omega)$, so (after extracting a subsequence, not relabeled) $u_k \rightharpoonup v$ weakly in $H^2(\Omega)$ for some $v \in H^2(\Omega)$. Since $u_k \to u$ in $L^\infty$, in particular $u_k \to u$ in $L^2$; passing to the limit in $L^2$ shows $v = u$ in $L^2$, hence $u \in H^2(\Omega)$. By weak lower semicontinuity,
\[
\|u\|_{H^2} \le \liminf_{k \to \infty} \|u_k\|_{H^2} \le C.
\]
Moreover,
\[
\int_\Omega u = \lim_{k \to \infty} \int_\Omega u_k = 0
\]
since $u_k \to u$ in $L^1$ (bounded domain and $L^\infty$–convergence). Thus $u \in \mathcal{B}_C$. Hence $\mathcal{B}_C$ is closed in $L^\infty$. Relative compactness with closedness in $L^\infty$, implies that $\mathcal{B}_C$ is compact in $L^\infty(\Omega)$. 
\end{proof}

	\begin{lemma}\label{lem:Q_continuous}
		Equip $\Theta=\mathcal B\times\mathcal H$ with the metric
		\(d\bigl((\bm \beta,h),(\tilde{\bm \beta},\tilde h)\bigr)=
		\|\bm \beta-\tilde{\bm\beta}\|+\|h-\tilde h\|_{\infty}\).
		Under Assumptions~\ref{assumption:identifiability}–\ref{assumption:density_P},
		the mapping
		\[
		Q(\theta)=\mathbb E\!\Bigl[
		\delta\bigl(\bm X^{\!\top}\bm \beta
		+h(\bm P)
		-\log s^{(0)}(\bm \beta,h,Y)\bigr)
		\Bigr],
		\qquad \theta=(\bm \beta,h)\in\Theta,
		\]
		is continuous on \(\Theta\).
	\end{lemma}
     \begin{proof}
    Write $s^{(0)}(\bm\beta,h,y)=\mathbb E\!\left[I(Y\ge y)\exp\{\bm X^{\!\top}\bm\beta+h(\bm P)\}\right]$.
    By Assumption~\ref{assumption:density_P}(i) there is $M_{\mathcal X}<\infty$ with
    $\|\bm X\|\le M_{\mathcal X}$ a.s., and by the definition of $\mathcal B$ we have
    $\|\bm\beta\|\le M_{\mathcal B}$. Moreover, by Sobolev embedding in $d=2$
    applied to $\mathcal H$ (bounded in $H^2$), there exists $M_\infty<\infty$ such that
    $\|h\|_\infty\le M_\infty$ for all $h\in\mathcal H$. Set
    \[
    C_*=M_{\mathcal B}M_{\mathcal X}+M_\infty.
    \]
    Then for all $(\bm\beta,h)\in\Theta$ and all $y\ge 0$,
    \begin{equation}\label{eq:s-bounds}
    e^{-C_*}\,\mathbb P(Y\ge y)\ \le\ s^{(0)}(\bm\beta,h,y)\ \le\ e^{C_*}\,\mathbb P(Y\ge y),
    \end{equation}
    since $e^{\bm X^\top\bm\beta+h(\bm P)}\in[e^{-C_*},e^{C_*}]$ a.s.  In particular,
    $s^{(0)}(\bm\beta,h,Y)>0$ a.s.
    
    Fix two parameters $\theta_1=(\bm\beta_1,h_1)$ and $\theta_2=(\bm\beta_2,h_2)$.
    By the mean–value theorem for the exponential and the bound above,
    \[
    \begin{aligned}
    \big|s^{(0)}(\theta_1,y)-s^{(0)}(\theta_2,y)\big|
    &= \left|\mathbb E\!\left[I(Y\ge y)\big(e^{\eta_1}-e^{\eta_2}\big)\right]\right| \\
    &\le \mathbb E\!\left[I(Y\ge y)\,e^{\max\{\eta_1,\eta_2\}}
    \left|(\bm X^\top(\bm\beta_1-\bm\beta_2))+(h_1-h_2)(\bm P)\right|\right] \\
    &\le e^{C_*}\,\mathbb P(Y\ge y)\,\big(M_{\mathcal X}\|\bm\beta_1-\bm\beta_2\|
    +\|h_1-h_2\|_\infty\big),
    \end{aligned}
    \]
    where $\eta_j=\bm X^\top\bm\beta_j+h_j(\bm P)$.
    Using \eqref{eq:s-bounds} and the mean–value theorem for $\log$,
    \begin{align*}
    \big|\log s^{(0)}(\theta_1,y)-\log s^{(0)}(\theta_2,y)\big|
    \ &\le\ \frac{\big|s^{(0)}(\theta_1,y)-s^{(0)}(\theta_2,y)\big|}
    {\min\{s^{(0)}(\theta_1,y),s^{(0)}(\theta_2,y)\}}\\
    \ &\le\ e^{2C_*}\,\big(M_{\mathcal X}\|\bm\beta_1-\bm\beta_2\|
    +\|h_1-h_2\|_\infty\big),
    \end{align*}
    where the factor $\mathbb P(Y\ge y)$ cancels, so the bound is uniform in $y$.
    
    Define, for $z=(Y,\delta,\bm X,\bm P)$ and $\theta=(\bm\beta,h)$,
    \[
    \zeta(z;\theta)=\delta\Big(\bm X^\top\bm\beta+h(\bm P)-\log s^{(0)}(\bm\beta,h,Y)\Big).
    \]
    Then for any $\theta_1,\theta_2\in\Theta$,
    \[
    \begin{aligned}
    \big|\zeta(z;\theta_1)-\zeta(z;\theta_2)\big|
    &\le \delta\Big(\big|\bm X^\top(\bm\beta_1-\bm\beta_2)\big|+\big|(h_1-h_2)(\bm P)\big|
    +\big|\log s^{(0)}(\theta_1,Y)-\log s^{(0)}(\theta_2,Y)\big|\Big)\\
    &\le \delta\,L_*\Big(\|\bm\beta_1-\bm\beta_2\|+\|h_1-h_2\|_\infty\Big),
    \end{aligned}
    \]
    with $L_*=(1+e^{2C_*})\max\{M_{\mathcal X},1\}$, which is deterministic and finite.
    Taking expectations and using $\delta\le 1$,
    \[
    |Q(\theta_1)-Q(\theta_2)|
    \le \mathbb E\big[\,|\zeta(z;\theta_1)-\zeta(z;\theta_2)|\,\big]
    \le L_*\,d(\theta_1,\theta_2).
    \]
    Thus $Q$ is globally Lipschitz on $(\Theta,d)$.
    \end{proof}

    \begin{lemma}\label{lem:Q_C2}
    Consider the metric space $(\Theta,d)$ of Lemma~\ref{lem:Q_continuous} and
    retain Assumptions~\ref{assumption:identifiability}--\ref{assumption:density_P}.
    Let $\theta=(\bm\beta,h)\in\Theta$, and directions
    $\eta_1=(\bm u_1,g_1)$, $\eta_2=(\bm u_2,g_2)$ in the linear space
    $\mathbb R^b\times\{g:\|g\|_\infty<\infty,\ \int_\Omega g=0\}$ such that
    $\theta+t\eta_r\in\Theta$ for all sufficiently small $t$ (e.g., $\theta$ is in the
    relative interior of $\Theta$).
    For $k=0,1,2$ define
    \[
    s^{(k)}(\theta,y)[\eta_1,\ldots,\eta_k]
    =\E\!\Bigl[
    \mathbf 1\{Y'\!\ge y\}\,
    \exp\bigl(\bm X'^{\!\top}\bm\beta+h(\bm P')\bigr)\,
    \prod_{r=1}^k\bigl(\bm X'^{\!\top}\bm u_r+g_r(\bm P')\bigr)
    \Bigr],
    \]
    where $(Y',\bm X',\bm P')$ is an i.i.d.\ copy of $(Y,\bm X,\bm P)$ independent of
    $(Y,\bm X,\bm P)$.
    Then $Q:\Theta\to\mathbb R$ is twice Fr\'echet differentiable on the relative
    interior of $\Theta$, with
    \[
    DQ(\theta)[\eta_1]
    =\E\!\Bigl[\,
    \delta\Bigl(
    \bm X^{\!\top}\bm u_1+g_1(\bm P)
    -\frac{s^{(1)}(\theta,Y)[\eta_1]}{s^{(0)}(\theta,Y)}
    \Bigr)
    \Bigr],
    \]
    \[
    D^2Q(\theta)[\eta_1,\eta_2]
    =\E\!\Bigl[\,
    -\delta\,
    \frac{
    s^{(2)}(\theta,Y)[\eta_1,\eta_2]\,
    s^{(0)}(\theta,Y)
    -s^{(1)}(\theta,Y)[\eta_1]\,
    s^{(1)}(\theta,Y)[\eta_2]
    }{
    \bigl(s^{(0)}(\theta,Y)\bigr)^2
    }
    \Bigr].
    \]
    \end{lemma}

    Assumption~\ref{assumption:density_P}~(i) gives $\|\bm X\|\le M_{\mathcal X}$ a.s.,
    $\mathcal B$ is compact so $\|\bm\beta\|\le M_{\mathcal B}$, and the
    Sobolev--Morrey embedding yields a uniform bound $\|h\|_\infty\le C_H M_{\mathcal H}$
    for all $h\in\mathcal H$. Let $C_*=M_{\mathcal B}M_{\mathcal X}+C_HM_{\mathcal H}$.
    Then for any $\theta=(\bm\beta,h)$ and $y\ge0$,
    \begin{equation}\label{eq:sbounds}
    e^{-C_*}\,\mathbb P(Y'\!\ge y)\ \le\ s^{(0)}(\theta,y)\ \le\
    e^{C_*}\,\mathbb P(Y'\!\ge y).
    \end{equation}
    Moreover, for $\eta=(\bm u,g)$ write $L(\eta)=M_{\mathcal X}\|\bm u\|+\|g\|_\infty$.
    By the mean–value theorem and boundedness of the exponential,
    \begin{align*}
    |s^{(1)}(\theta,y)[\eta]|
    &\le e^{C_*}\,\mathbb P(Y'\!\ge y)\,L(\eta),\\
    |s^{(2)}(\theta,y)[\eta_1,\eta_2]|
    &\le e^{C_*}\,\mathbb P(Y'\!\ge y)\,L(\eta_1)L(\eta_2).
    \end{align*}
    Combining with \eqref{eq:sbounds} yields the uniform-in-$y$ ratio bounds
    \begin{equation}\label{eq:ratio-bounds}
    \left|\frac{s^{(1)}(\theta,y)[\eta]}{s^{(0)}(\theta,y)}\right|
    \le e^{2C_*}L(\eta),\qquad
    \left|\frac{s^{(2)}(\theta,y)[\eta_1,\eta_2]}{s^{(0)}(\theta,y)}\right|
    \le e^{2C_*}L(\eta_1)L(\eta_2),
    \end{equation}
    and likewise
    $\bigl|s^{(1)}(\theta,y)[\eta_1]\,s^{(1)}(\theta,y)[\eta_2]/(s^{(0)}(\theta,y))^2\bigr|
    \le e^{4C_*}L(\eta_1)L(\eta_2)$.
    Now, for $t\mapsto\theta+t\eta_1$ define
    \[
    \Psi_t(Y,\delta,\bm X,\bm P)
    =\bm X^{\!\top}\bm u_1+g_1(\bm P)
    -\frac{s^{(1)}(\theta+t\eta_1,Y)[\eta_1]}{s^{(0)}(\theta+t\eta_1,Y)}.
    \]
    By dominated convergence, we have
    $\Psi_t\to \Psi_0$ a.s., and the uniform ratio bound \eqref{eq:ratio-bounds}
    gives $|\Psi_t|\le C(1+L(\eta_1))$ for a deterministic $C$.
    Thus
    \[
    \frac{Q(\theta+t\eta_1)-Q(\theta)}{t}
    =\E\bigl[\delta\,\Psi_t\bigr]\ \longrightarrow\ \E\bigl[\delta\,\Psi_0\bigr],
    \]
    which is the stated formula for $DQ(\theta)[\eta_1]$.
    
    Similarly, consider $t\mapsto DQ(\theta+t\eta_2)[\eta_1]$.
    By the chain rule for $\log$, Taylor’s formula, and the definitions of $s^{(k)}$,
    \[
    \frac{d}{dt}\Big|_{t=0}\left(\frac{s^{(1)}(\theta+t\eta_2,y)[\eta_1]}{s^{(0)}(\theta+t\eta_2,y)}\right)
    =\frac{s^{(2)}(\theta,y)[\eta_1,\eta_2]\,s^{(0)}(\theta,y)
    -s^{(1)}(\theta,y)[\eta_1]\,s^{(1)}(\theta,y)[\eta_2]}{(s^{(0)}(\theta,y))^2}.
    \]
    The interchange of differentiation and expectation that defines $s^{(k)}$ is
    justified by dominated convergence using the same $e^{C_*}$ and $L(\cdot)$ bounds,
    and the ratio is bounded uniformly in $y$ by \eqref{eq:ratio-bounds}.
    Therefore
    \begin{align*}
    &\frac{DQ(\theta+t\eta_2)[\eta_1]-DQ(\theta)[\eta_1]}{t}\\
    &\longrightarrow\
    -\E\!\left[\delta\,
    \frac{s^{(2)}(\theta,Y)[\eta_1,\eta_2]\,s^{(0)}(\theta,Y)
    -s^{(1)}(\theta,Y)[\eta_1]\,s^{(1)}(\theta,Y)[\eta_2]}
    {(s^{(0)}(\theta,Y))^2}\right].
    \end{align*}
    Again, the integrand is dominated by a deterministic constant times
    $L(\eta_1)L(\eta_2)$, so the limit passes through the expectation.
    Bilinearity and continuity in $(\eta_1,\eta_2)$ are immediate from the bounds,
    which also give the $o(\|\eta_2\|)$ remainder uniformly in $\eta_1$.
    Thus $Q$ is $C^2$ Fr\'echet on the interior of $\Theta$ with the
    stated $DQ$ and $D^2Q$.

	
	
	\begin{lemma}\label{lem:uniq_max}
		$Q(\theta)$ attains its unique global maximum in $(\Theta,d)$  at
		$\theta_0=(\beta_0,h_0)$.
	\end{lemma}
	
    \begin{proof}
        By Lemma~\ref{lem:Q_continuous}, $Q$ is continuous on $(\Theta,d)$ with
    $d\big((\bm\beta,h),(\tilde{\bm\beta},\tilde h)\big)
    =\|\bm\beta-\tilde{\bm\beta}\|+\|h-\tilde h\|_\infty$.
    By compactness of $\mathcal B$ (Assumption~\ref{assumption:identifiability}~(iii))
    and the compact embedding $H^2(\Omega)\hookrightarrow C^0(\Omega)$ in $d=2$,
    the set $\mathcal H=\{h\in H^2_{\mathbf n}(\Omega): \int_\Omega h=0,\ \|h\|_{H^2}\le M_{\mathcal H}\}$
    is compact in the $\|\cdot\|_\infty$–topology (see Lemma~\ref{lemma:compactness}).
    Hence $\Theta=\mathcal B\times\mathcal H$ is compact under $d$, and by Weierstrass
    $Q$ attains a global maximum on $\Theta$. Uniqueness follows from strict concavity of $Q$, see \eqref{eq:cocave}. The fact that the minimum is $\theta_0$ is implied by Theorem~\ref{theo:identification}.
    \end{proof}



\end{document}